\documentclass[onecolumn,usenatbib]{mn2e}
\usepackage{journals}
\usepackage{epsf}
\author[D. Johnston]{David E. Johnston,$^{1,2}$\\
$^1$Department of Physics, The University of
Chicago, 5640 South Ellis Avenue, Chicago, IL 60637\\
$^2$Princeton University Observatory, Peyton Hall, Princeton, NJ 08544}

\title{Measuring the Galaxy-Galaxy-Mass Three-point Correlation Function
with Weak Gravitational Lensing}

\begin{document}
\maketitle

\begin{abstract}
We discuss the galaxy-galaxy-mass three-point correlation function and show
how to measure it with weak gravitational lensing. 
The method entails choosing a large of pairs of
foreground lens galaxies and
constructing a mean shear map with respect to their axis,
by averaging the ellipticities of background source galaxies. 
An average mass map can be reconstructed from this shear map
and this will represent the average mass distribution around pairs of galaxies.
We show how this mass map is related to the projected
galaxy-galaxy-mass three-point correlation function. Using a large
N-body dark matter simulation populated 
with galaxies using the Halo Occupation Distribution (HOD) bias
prescription, we compute these correlation functions, mass maps, and 
shear maps. The resultant mass maps are
distinctly bimodal, tracing the galaxy centers and remaining 
anisotropic up to scales much larger than the galaxy separation.
At larger scales, the shear is approximately
tangential about the center of the pair but with small azimuthal variation
in amplitude. We estimate the signal-to-noise ratio of the reconstructed 
mass maps for a survey of similar depth to the Sloan Digital Sky
Survey (SDSS) and conclude that the galaxy-galaxy-mass three-point function 
should be measurable with the current SDSS weak lensing data. 
Measurements of this three-point function, along with galaxy-galaxy lensing
and galaxy auto-correlation functions, will provide new constraints 
on galaxy bias models. The anisotropic shear profile around 
close pairs of galaxies is a prediction of cold dark matter models 
and may be difficult to reconcile with alternative theories of gravity 
without dark matter.
\end{abstract}

\begin{keywords}
gravitational lensing
\end{keywords}

\section{Introduction}

The last few years have seen the emergence of an observationally 
consistent cosmological model based on firm physical ideas.
The hot big bang model of the expanding universe, whose structure and 
dynamics can be described by general relativity (GR), has a number of 
cosmological parameters that are now well constrained by numerous
observations. WMAP measurements of the cosmic 
microwave background radiation (CMB) 
\citep{spergel:wmap-parameters}, supernova measurements
\citep{riess:highz-supernova,perlmutter:supernova-expansion}, and measurements of
large scale structure \citep{tegmark:pk} among others, have settled on a 
concordance model of a spatially flat universe with matter density about 25\%
of critical. This concordance model is consistent with the ages of the oldest stars
\citep{jimenez:oldest-stars}, Hubble parameter measurements 
\citep{freedman:hubble-constant}, and primordial abundances of light elements
from Big Bang Nucleosynthesis (BBN) \citep{burles:bbn}.

One of the main requirements of a successful cosmological model 
is that it connect measurements
of the early universe, e.g. the CMB and BBN, to the measurements of the local universe
of galaxies and stars. The large scale structure of galaxies that we observe
in galaxy surveys must have formed through gravitational collapse of the small
fluctuations left over from that earlier time. The properties of these
large-scale mass densities can be predicted from the initial conditions as
observed in the CMB combined with our understanding of gravity as described by
general relativity. However, since the Universe in our model is dominated 
by dark matter, we need to make the connection between the distribution of 
galaxies relative to the dark matter if we want to compare these observed 
structures to what is predicted.

The relation between the galaxy density and the dark matter density is known as
the \emph{bias}. For the concordance model to be in agreement 
with the data on galaxy clustering, it
is required that on large scales the bias parameter
(which we will define shortly) is close to one. On smaller scales the bias must
be different from one but of order unity. Moreover, the relative bias 
of different galaxy types is known to depend on spectral type and luminosity. 
One can do better than simply treating the bias as a free parameter. Galaxy
biasing is after all closely linked to the physics of galaxy formation. Although
galaxy formation is not completely understood, it is becoming increasingly
amenable to theoretical modeling and numerical simulation. Gas dynamical N-body
simulations and Semi-analytic models of galaxy formation 
\citep{benson:sams,kauffmann:sams} each predict 
how galaxies are biased with respect to mass. 
Nevertheless, sufficient uncertainty on the details of galaxy formation remain
that observational constraints on the bias are needed.

The apparent complexity of the bias has motivated the development
of new probes of large-scale structure and bias that do not rely on luminous
galaxies as tracers of the underlying mass. Gravitational lensing is a prime
example. Lensing is ideally suited to studying dark matter since it is directly 
sensitive to the total mass density along the line of sight.
Other probes of dark matter such as dynamical measurements must rely 
on some assumptions such as that a system has reached
dynamical equilibrium. By contrast, lensing provides a clean probe of the 
dark matter. 

Galaxy-galaxy lensing (GGL), in particular, probes the dark matter
density profile around luminous galaxies. In galaxy-galaxy lensing, one
measures the mean tangential distortion of a large number of source galaxies
behind foreground lens galaxies by stacking the signal from the latter
\citep{sheldon:gmcf}. To date, lensing measurements and theory have 
focused on the two-point galaxy-mass correlation function (2PCF). This correlation 
function can be measured from galaxy photometric surveys and compared to the 
results of N-body dark matter simulations. These measurements provide 
direct constraints on bias models. Although they do not
measure the galaxy bias parameter, $b$ , directly, they do measure a
closely related quantity and so these measurements do test the bias model.

While the 2PCF is the lowest order 
measurement of how much a distribution is clustered, it does not 
carry complete statistical information about a non-Gaussian
distribution. The rest of the information
is encoded in the infinite set of N-point correlation functions.
In particular the higher order ($\mbox{N} > 2$) functions provide the 
information about the coherence and shapes of large-scale 
structures such as filaments, walls, and voids
that are apparent in both data and simulations. These higher order
correlation functions add further constraints to be met by 
cold dark matter (CDM) models and new probes of the bias.

For the most part observational studies of higher 
order correlation functions have measured correlations between galaxies.
In this paper we will describe a method for measuring the galaxy-galaxy-mass 
three-point correlation function (GGM3PCF) with weak lensing.
This method is a generalization of the method for measuring 
the galaxy-mass two-point correlation function (GM2PCF) 
with galaxy-galaxy lensing. Just as the GM2PCF provided new constraints on
the bias model, so too does the GGM3PCF.
While the GM2PCF is a measure of the isotropic mass density around galaxies,
the GGM3PCF gives us information 
about the shape of the mass distribution around pairs of galaxies.
Together with galaxy N-point functions 
and the GM2PCF, the GGM3PCF provides independent constraints on models
of galaxy biasing that are crucial for understanding both large scale
structure and the processes of galaxy formation.

It is reasonable to assume that
CDM models should predict that the mass around a pair
of galaxies will have an anisotropic distribution. A priori it may be 
purely elliptical or clumped tightly about each galaxy with a 
distinctly bimodal appearance. This expectation comes from the
fact that halos are not themeselves spherically symmetic and 
that the galaxies should also trace this ellipticity. Also the
galaxies themselevs have mass associates with them; stellar
mass at the least and probably dark matter sub-halos.
Whatever the details of the distribution is, it will create a 
gravitational lensing effect that is also
anisotropic. This anisotropic lensing signal should be measurable
with current and future weak lensing surveys.

Another use of the GGM3PCF is to test for the existence
of dark matter itself. Alternative theories of gravity with no dark
matter, e.g. MOND \citep{milgrom:mond}, must eventually make 
prediction for gravitational lensing.
This predicted lensing profile must agree with the lensing observations.
With no dark matter, the lensing effect is 
only caused by the visible matter. Two galaxies that are 
close together will simply look like a monopole at large distance 
from them and the lensing effect produced should be nearly
isotropic regardless of the form of the lensing profile.
However, dark matter models predict (as we will 
demonstrate) that galaxy pairs have an
extended dark matter halo about them that is anisotropic even at
large distance from their center. This causes an anisotropic
lensing effect at these large distances. Thus, dark matter models
and alternative models should make different predictions
for the lensing anisotropy and thus test the idea of the existence of dark 
matter. Hoekstra et al. 2003 have already 
claimed to have measured halo flattening around single 
galaxies, a similar effect, with weak lensing and argue that it 
would be difficult to reconcile this anisotropy with theories 
such as MOND. Measurements of the GGM3PCF
would allow one to extend this result to scales larger than the size of
individual galaxies. Also, since the GGM3PCF deals with only the locations of galaxies,
it would not have to rely on simulations that predict how the shape of 
the galaxy light correlates with the shape of the halo.

In this paper, we show how the lensing pattern around pairs of galaxies is
related to the GGM3PCF. In particular, we show how
the 2D projection of the 3D GGM3PCF is directly
measurable with weak lensing measurements such as the measurements being conducted
\citep{sheldon:gmcf,mckay:gal-gal} with data from the
Sloan Digital Sky Survey \citep{york-short:sdss}.
We use an N-body dark matter simulation populated with galaxies using 
a bias model chosen to reproduce the results of the hydrodynamical
simulations and make numerical calculations of the
various two and three-point correlation functions in both 2D and 3D.
We calculate the lensing effect and estimate the amount of noise
that should be present to study the signal-to-noise that can be obtained.
We conclude that the GGM3PCF can now be measured
with the present SDSS data.

\section{Correlation Functions and their Applications to Cosmology}
\label{section:corr}

We begin by reviewing galaxy correlation functions and then describe their
extension to galaxy-mass cross-correlations. 
The probability of finding a galaxy (or any mass point) in an infinitesimal volume 
of space $dV_1$ is simply
$dP_{g} = \rho_g dV_1$ where $\rho_g$ is the mean galaxy density.
The probability of finding one galaxy in $dV_1$ \emph{and} another in
$dV_2$ defines the two-point correlation function $\xi_{gg}(r)$ 
\[ dP_{gg} = \rho_g^2 dV_1 dV_2 ~ [ 1 + \xi_{gg}(r) ] \]   
The function $\xi_{gg}(r)$ can be thought of as the contribution of
clustering to the joint probability in excess of random. By the 
assumed large scale homogeneity and isotropy, $\xi_{gg}$ can depend only
on $r$ and not on angle or location in space. The marginal probability of
finding a galaxy in $dV_2$ given that one already has found one in $dV_1$
is $dP(g_2|g_1)= \rho_g^2 dV_1 dV_2 ~ [ 1 + \xi_{gg}(r) ]/\rho_g dV_1 = \rho_g dV_2 ~ [ 1 + \xi_{gg}(r) ]$.

We can further consider the probability of finding a third galaxy
in infinitesimal volume $dV_3$ in addition to the two in $dV_1$ and $dV_2$.
The resulting joint probability defines the three-point 
correlation function (3PCF),  $\zeta_{ggg}$ ,

\[ dP_{ggg} = \rho_g^3 dV_1 dV_2 dV_3~ \left[ ~1 + \xi_{gg}(r_{12}) 
+ \xi_{gg}(r_{13}) + \xi_{gg}(r_{23}) +\zeta_{ggg} \right] \] 

Here $r_{ij}$
refers to the separation between galaxies $i$ and $j$. The point of defining
$\zeta_{ggg}$ so as to result in a 5-term expression is similar to the reason we defined
a 2-term expression for the 2-point case. 
If the third galaxy were simply put down at random or not clustered
with the first two, or very far from the first two, we would 
have $ dP_{ggg} = \rho_g^3 dV_1 dV_2 dV_3 [1 + \xi_{gg}(r_{12})]$.
Thus the other three terms, $\xi_{gg}(r_{13})$, $\xi_{gg}(r_{23})$, and $\zeta_{ggg}$,
only contribute the \emph{excess} over this random occurrence
and by symmetry of this argument we also get the next two terms $\xi_{gg}(r_{13})$
and $\xi_{gg}(r_{23})$ . The last term $\zeta_{ggg}$
is needed since there is no reason to assume that the first four terms 
completely describe the joint probability. 

The volume averages of the 2PCF, 3PCF, and higher order
correlation functions are the moments of the
probability distribution of the smoothed density field, e.g. the
variance, skewness, kurtosis etc. Since a Gaussian distribution
is characterized just by its 2PCF, the 3PCF is the lowest-order
N-point statistic that characterizes non-Gaussianity of the density
field. Generally non-linear gravitational effects generate a non-Gaussian
density distribution even if the initial density field is Gaussian.

One can write down expressions for the joint probabilities
of still more galaxies and these in turn will define the N-point correlation 
functions. The 3PCF $\zeta_{ggg}$, due to the large scale homogeneity and isotropy, 
is only a function of the three triangle sides
$r_{12}$,$r_{13}$, $r_{23}$ (or any other parametrization of the 
triangle).

One can also define cross correlation functions between different species
of objects, for example, between galaxies and dark matter 
mass particles. We can write down joint probabilities of finding 
a galaxy and a mass particle, a galaxy
and two mass particles, and two galaxies and a mass particle respectively 
(taking all the $dV$ to be equal in size for simplicity),

\begin{eqnarray}
 dP_{gm} = & \rho_g \rho_m dV^2   & [1 + \xi_{gm}(r)] \\
 dP_{gmm} = & \rho_g \rho_m^2 dV^3 & [1 + \xi_{gm}(r_{g,m_1}) + \xi_{gm}(r_{g,m_2}) + \xi_{mm}(r_{m_1,m_2}) +\zeta_{gmm}]  \\
 dP_{ggm} = & \rho_g^2 \rho_m dV^3 & [1 + \xi_{gg}(r_{g_1,g_2}) + \xi_{gm}(r_{g_1,m}) + \xi_{gm}(r_{g_2,m}) +\zeta_{ggm}] \label{eq:ggm3}
\end{eqnarray}

These joint probabilities define the three two-point 
functions, $\xi_{gg}$ , $\xi_{gm}$, $\xi_{mm}$ and the
four three-point functions $\zeta_{ggg}$ , $\zeta_{ggm}$, $\zeta_{gmm}$ and $\zeta_{mmm}$.

One can also make the same definitions in 2D for a set
of points that have been projected over the third dimension. In this 
case we will refer to these 2D correlation functions as, for example, 
$w_{gm}$ for the two-point and $z_{ggm}$ for the three-point functions.


The uses of correlation functions in cosmology are many. 
In the inflation scenario, structure formation is seeded by random
quantum fluctuations. In most models these initial fluctuations 
are predicted to be Gaussian
distributed and therefore completely described by their
two-point correlation $\xi(r)$ or its Fourier transform 
the power spectrum $P(k)$. In this scenario, the 3PCF
and higher order correlation functions that arise are
entirely due to the onset of  non-linear gravitational collapse and so should be predictable with 
dark matter simulations given a set of cosmological parameters
and a galaxy bias prescription.

The first measurements of the galaxy two-point correlation function 
\citep{totsuji:2pcf,peebles:2pcf,groth-peebles:2pcf,gott-turner:2pcf}
found that it was well described by a power law,
$ \xi_{gg}(r) = (r/r_0)^{-\gamma}$, with $\gamma=1.8$ and $r_0 \approx 5 h^{-1}$ Mpc.
The most recent measurements \citep{zehavi:departure} are now
sensitive enough to detect small deviations from a power law, in agreement
with predictions that use N-body simulation combined with HOD bias 
models (see Section \ref{section:hod}). 

The first measurement of the 3PCF for galaxies was by \cite{peebles-groth:3pcf}. 
For these early correlation function measurements,
there was no redshift information, so these were measured in
terms of angular coordinates $\theta_{12}$ rather than physical
length scales $r_{12}$. Constraints on the 3D 2PCF and 3PCF
came from inverting the projections. From this first
measurement, it was found that to experimental accuracy
the 3PCF could be written in a reduced form relating it to the
2PCFs (on scales less than a few Mpcs) as 
\begin{equation}
 z_{ggg} =~Q~ [
  w_{gg}(\theta_{12}) w_{gg}(\theta_{13}) 
+ w_{gg}(\theta_{12}) w_{gg}(\theta_{23}) 
+ w_{gg}(\theta_{13}) w_{gg}(\theta_{23}) ] \label{eq:zggg}
\end{equation}
with Q being a constant $\approx 1$. This scaling relation is
referred to as  hierarchical scaling.
Although much effort has gone into
deriving such an equation from first principles, no one has shown
analytically why this form should be true. However N-body
simulations do indicate that the 3PCF approximately takes this form
in the non-linear regime probed by these early measurements.
On larger scales ($r > 10 h^{-1}$ Mpcs) where one can treat
gravity perturbatively, this factor Q is indeed independent
of scale however there is a non-vanishing dependence on triangle shape
\citep{fry:3pcf-hierarchy}. 
This large scale shape dependence is a quantitative statement that
co-linear triangles or filamentary structures are more likely than
isosceles triangles or isotropic structures and 
physically relects the fact that 
gravitational collapse happens anisotropically with velocity
flows along density gradients \citep{bernardeau:lss-pert}. 
A visual inspection of recent 
N-body simulations \citep{jenkins:virgo,evrard:hubble-volume} 
as well as real data from galaxy redshift surveys 
\citep{huchra-geller:cfa,peacock:2df,tegmark:pk}
show the existence of large-scale filaments on the outskirts 
of giants voids.

The basic hierarchical scaling of Equation \ref{eq:zggg}
was confirmed by \cite{fry-seldner:bispectrum}
however these early measurements from the Lick galaxy catalog did 
not extend to large enough scales to reveal
the expected configuration-dependent signature 
from gravitational instability \citep{fry:3pcf-hierarchy}.
Later results by \cite{frieman:3pt-APM} using the APM survey data,
\cite{scoccimarro:bispectrum} using the IRAS 1.2 Jy survey
and \cite{feldman:bispectrum} using 
the IRAS PSCz survey were in good agreement with the predictions from
the gravitational instability picture and provided new constraints 
on the bias for these samples.

These studies have had much success at measuring the 2PCF and 3PCF
of galaxies. Furthermore N-body simulations have been able to make
precise predictions, given the cosmological parameters, of the 2PCF and 3PCF
of the mass density. However until recently the galaxy-mass correlation
functions have remained unmeasured. The measurements of the 3PCF of galaxies
in fact can be used to constrain the large scale bias between galaxies
and mass but the smaller scale bias has mostly remained a mystery.
The fact that on small scales the observed 2PCF of galaxies and the
2PCF of mass from N-body simulations disagree leads to the requirement that
there must be scale dependent bias between galaxies and mass
if the CDM framework is correct.
Another indicator of this complexity is that different populations 
of galaxies have different measured correlation 
functions and therefore are biased differently 
relative to each other. There is no reason why 
any particular sample of galaxies should trace the dark matter exactly.
Trying to measure and model the bias using galaxy-mass correlations 
is one of the main goals of galaxy-galaxy lensing and the galaxy-mass
3PCF should help achieve this goal.

\section{Measuring Correlation Functions with Weak Lensing} \label{section-weak}

The general relativistic lens equation relates the bending angle of
light rays to the mass distribution. In the thin lens approximation
the deflection can be computed from the projected surface mass density
$\Sigma(x,y) = \int dz \rho(x,y,z) $ where $z$ denotes the coordinate along
the line of sight. It is customary make this dimensionless
by dividing by the geometric factor
$ \Sigma_{crit} = c^2 D_S / 4 \pi G D_L D_{LS} $,
where the distances are the angular diameter distances to the lens, 
to the source, and from the lens to the source. 
This defines a dimensionless 2D mass density, or convergence, 
$\kappa(x,y)=\Sigma(x,y)/\Sigma_{crit}$. It is useful to define a 2D potential 
\begin{equation}
\phi(\vec{x}) = \frac{1}{\pi} \int d^2x^{\prime} \kappa(\vec{x}^{\prime}) 
\ln |\vec{x} -\vec{x}^{\prime}|  
\label{eq:lens-potential}
\end{equation}
One can show that $\kappa$ and $\phi$ satisfy Poisson's equation 
\begin{equation}
 2 \kappa= \nabla^2 \phi = \phi_{,11} + \phi_{,22} \label{eq:kappa-poisson}
\end{equation}
where the subscripts with commas denote derivatives transverse to the line of sight. 
The other second derivatives of $\phi(\vec{x})$ define the shear,
\begin{equation}
2 \gamma_1  \equiv  \phi_{,11}-\phi_{,22} ~~~ \mbox{and} ~~~
 \gamma_2   \equiv  \phi_{,12}	 \label{eq:def-shear}
\end{equation}
The two-component polar $\gamma$ is related to 
the anisotropic stretching of the source galaxy
shapes in a linear way. With the ellipticity components defined 
in terms of the second moments of the galaxy light 
then it can be shown that $\langle e_i \rangle = 2 \gamma_i$. Thus measurements
of galaxy ellipticities provide an estimate of the shear.
These measurements are noisy because galaxies are intrinsically elliptical
with $\langle e_i^2 \rangle ^{1/2} \approx 0.3$. Therefore accurate shear
measurements require one to average over enough galaxies to reduce this
``shape noise'' to $(\langle e_i^2 \rangle /N )^{1/2}$.
See \cite{bartelmann-schneider:weak-review} for a review. 

Because $\kappa$, $\gamma_1$, and $\gamma_2$ 
are all related to $\phi$ it is possible to determine
$\kappa$ from the measurable $\gamma$. One such relation is 
\begin{equation} 
\nabla \kappa = \left (  \begin{array}{c}
	\gamma_{1,1} + \gamma_{2,1} \\ \gamma_{2,1} - \gamma_{1,2} \label{eq:kaiser-del}
	\end{array} \right )    
\end{equation}
To solve for $\kappa$ one needs to solve this partial differential
equation, which states that the derivative of $\kappa$ is locally
related to the derivative of the shear. By Fourier transforming this equation
(or Equations \ref{eq:def-shear} and \ref{eq:kappa-poisson}) one can see that
$\kappa$ is locally related to the shear in Fourier space, but in real
space it is related to the shear non-locally (i.e. it involves a convolution).
There is also the problem of the mass sheet degeneracy: $\kappa$
can only be measured up to a constant with shear data.

Weak lensing first became a useful tool in cosmology as a method of
measuring the projected mass around rich clusters 
of galaxies \citep{fahlman:cluster-lensing,clowe-luppino:cluster,
joffre:lensing-3667,dahle:weak-lensing-clusters,tyson-fischer:1689,luppino-kaiser:cluster}.
Rich clusters are massive enough that the surface mass density is close to the
critical surface mass density $\Sigma_{crit}$ (sometimes greater) 
and so there are sometimes
giant arcs \citep{hammer:arcs,kneib:abell2218} and even if not, 
the lensing shear is often very large and so can be measured reliably. 
Various methods of inverting the shear map to produce
2D mass maps for clusters have been developed 
\citep{lombardi-bertin:inversion,seitz-schneider:inversion},
based on the original work by \cite{kaiser-squires:inversion}. We will refer to any
of these algorithms simply as KS algorithms. 

\section{Galaxy-galaxy lensing}
\label{section:gal-gal-lensing}
Galaxy-galaxy lensing (GGL) denotes the lensing of background source 
galaxies by individual foreground 
lens galaxies. In this case, one does not have large enough signal-to-noise 
ratio to reliably measure the signal around each lens galaxy;
instead one must average the signal for many lens
galaxies and be content with this statistical measure. The first attempt to measure
GGL \citep{tyson:gal-gal} found only an upper limit.
The first detection was by \cite{brainerd:gal-gal} followed by
\cite{dellantonio:gal-gal,griffiths:gal-gal,hudson:gal-gal}. The first very
significant measurement was by \cite{fischer:gal-gal} in the SDSS 
followed by other high S/N results 
\citep{wilson:gal-gal,smith:gal-gal,mckay:gal-gal,hoekstra:gal-gal-cnoc}.
The latest result by \cite{sheldon:gmcf} makes high signal-to-noise ratio
measurements of the 3D galaxy-mass correlation function and the bias.

The early GGL studies concentrated more on small 
scales ($R \leq 1~h^{-1}$ Mpc) and were interested in
constraining the mass to light ratios by modeling the mass profiles as isothermal 
spheres and extracting a velocity dispersion. However the best way to interpret 
GGL, especially at larger scales, is as a measurement 
of the galaxy-mass two-point correlation function (GM2PCF),
$\xi_{gm}(r)$. The average density of mass around a galaxy can be written
\begin{equation} \rho_m(r)= \rho_{crit} \Omega_m [ 1+ \xi_{gm}(r) ] \end{equation}
After dropping the constant mass sheet which produces no lensing, the 2D mass over-density
is just the projection 
\begin{equation}
\Sigma(R) = \rho_{crit} ~ \Omega_m \int dz \xi_{gm}(r) 
\equiv \rho_{crit} ~ \Omega_m w_{gm}(R) \label{eq:proj-sigma}
\end{equation}
where $z$ denotes the radial coordinate; i.e., $r^2 = R^2 + z^2$.
This defines the projected GM2PCF, $w_{gm}(R)$, where
$R$ is the distance from galaxy to mass particle projected in the plane 
perpendicular to the line of sight. When the average mass profile is circularly 
symmetric, as in this case, the mean tangential shear $\gamma_T$ in an annulus 
of radius $R$ can be 
expressed in terms of $\Sigma(R)$ as 
\begin{equation} \Sigma_{crit} \gamma_T(R) \equiv \Delta \Sigma (R)
= \overline{\Sigma}(<R) - \Sigma(R) \end{equation}
where $\overline{\Sigma}(<R)$ denotes the average $\Sigma$ 
inside a circle of radius $R$ and $\Sigma(R)$ is the mean in a narrow
annulus of radius $R$.
Written out in terms of $w_{gm}$, this is
\begin{equation}
\Delta \Sigma(R) = \rho_{crit} ~ \Omega_m \left[ \frac{2}{R^2} 
\int_0^R dR^{\prime} R^{\prime} w_{gm}(R^{\prime})-w_{gm}(R) \right] \label{eq:int-wgm}
\end{equation}

Differentiating with respect to $R$, one arrives at
\begin{equation}
\rho_{crit} ~ \Omega_m \frac{d w_{gm}(R)}{dR}
= \frac{d \Delta \Sigma(R)}{dR} 
+ 2 \frac{\Delta \Sigma(R)}{R} \label{eq:deriv-wgm}
\end{equation}

One could then integrate this equation to get $w_{gm}(R)$ in terms of 
tangential shear measurements. The unknown constant 
of integration is simply due to the usual mass sheet degeneracy. This equation is the
circularly symmetric equivalent to Equation \ref{eq:kaiser-del} and in fact
can be derived by writing that equation in polar coordinates and assuming
$\kappa$ is only a function of $R$. However rather than integrating this
equation to get $w_{gm}(R)$ one can calculate the 3D $\xi_{gm}(r)$
directly. Since $\xi_{gm}(r)$ is spherically
symmetric, there exists an inversion formula for the projection called the
Abel integral formula \citep{binney-tremaine:gal}.
\begin{equation}
\xi_{gm}(r)=\frac{-1}{\pi} \int_r^\infty dR \frac{d w_{gm}(R)}{dR}
\frac{1}
{\sqrt{R^2-r^2}} \label{eq:xi-invert}
\end{equation}
This equation was also used by \cite{saunders:2pt} to obtain the
galaxy-galaxy auto-correlation function from the measured projected
correlation function in the IRAS survey.
Equation \ref{eq:deriv-wgm} can be substituted into Equation 
\ref{eq:xi-invert} to compute the galaxy 
mass correlation function from the shear measurements.
This inversion method was recently used in \cite{sheldon:gmcf} to 
make the first direct measurements of the 3D GM2PCF from weak lensing.

One thing to note about Equation \ref{eq:xi-invert} is that 
one does not have data out to infinitely large $R$. If one has data from 
scales $R_{min}$ to $R_{max}$ then  
one can measure $\xi_{gm}(r)$ only on these scales, and one can perform the
integration only up to $R_{max}$,
\begin{equation}
\xi_{gm}(r > R_{min})=\frac{-1}{\pi} \int_r^{R_{max}} dR 
\frac{d w_{gm}(R)}{dR} \frac{1}
{\sqrt{R^2-r^2}} + C(r) \label{eq:xi-invert-c}
\end{equation}
where $C(r)$ is a correction 
\begin{equation}
C(r)=\frac{-1}{\pi} \int_{R_{max}}^\infty dR 
\frac{d w_{gm}(R)}{dR}
\frac{1}{\sqrt{R^2-r^2}}
\end{equation}
To evaluate the correction term, one can either extrapolate $w_{gm}(R)$ 
beyond the data region or use a model for $\xi_{gm}(r)$. Either way, this
results in a model dependence or uncertainty in the correction. Fortunately, 
because $\xi_{gm}(r)$ drops off rapidly (typically like $r^{-1.8}$ or 
faster at large scales)
this correction term is negligible for any reasonable $\xi_{gm}(r)$ 
except for $r$ close to $R_{max}$. The first applicaion of this
inversion method to galaxy-galaxy lensing was \cite{sheldon:gmcf}
and a complete treatment of the method is given in \cite{johnston:inversion}.

\section{Measuring the higher order functions}
\label{section:measuring-higher}

Just as the galaxy-mass two-point correlation function can be measured with 
the lensing signal around a galaxy, the galaxy-galaxy-mass three-point
correlation function (GGM3PCF)
can be measured with the lensing signal around pairs of galaxies.
Here we derive the convergence field $\kappa$ around pairs of 
galaxies separated by a projected distance $R_{12}$.

In 3D, the probability of finding two galaxies and a mass particle in some
triangular configuration in three infinitesimal volume elements, 
each of common size $dV$, is given by Equation \ref{eq:ggm3}. We 
call this $dP_{ggm}(r_{12},r_{13},r_{23})$, where
1 and 2 indicate the galaxies and 3 indicates the mass particle.
When we already have a pair of galaxies and just ask what is the 
probability of finding a mass particle, this is the conditional 
probability
\begin{equation}
dP(m|g_1,g_2)=dP_{ggm}/dP_{gg}
\end{equation}
Dividing this expression by $dV$ and multiplying by the mass per particle 
yields the average \emph{density} of mass around pairs of galaxies 
separated by $r_{12}$. 

\begin{eqnarray}
\rho_m(r_{13},r_{23}|r_{12}) & = & \rho_{crit} \Omega_m~ 
\left[ ~ \frac{1 + \xi_{gg}(r_{12}) +\xi_{gm}(r_{13}) +\xi_{gm}(r_{23}) 
+\zeta_{ggm}}{1 + \xi_{gg}(r_{12})} ~\right] \nonumber \\
  & = &\rho_{crit} \Omega_m~\left[~ 1+ \frac{\xi_{gm}(r_{13}) +\xi_{gm}(r_{23}) 
+\zeta_{ggm}}{1 + \xi_{gg}(r_{12})} ~\right] \label{eq:rho-pair-3D}
\end{eqnarray}
With weak lensing, we find galaxy pairs at some projected distance
$R_{12}$ and want to know the 2D projected mass density around these galaxies.
So we need to project the mass over the radial direction but also average over
the unknown radial distance between the two galaxies. Even though we may have
measured redshifts for both galaxies, they cannot be used to calculate this 
radial distance precisely because the radial 
peculiar velocities cause redshift distortions.
Following \cite{peebles:lss}, where a similar projection is performed, 
we define $u$ to be the unknown radial distance 
between the two galaxies and $v$ to be the unknown radial distance between
galaxy 1 and the mass particle (labeled 3). The radial distance between galaxy
2 and the mass particle therefore is $(v-u)$, since the three points define
a triangle. With these two numbers $u$ and $v$ we can write the 3D distances in
terms of the 2D projected distances 
\begin{eqnarray}
r_{12}^2 & = & R_{12}^2 + u^2 \nonumber \\
r_{13}^2 & = & R_{13}^2 + v^2 \\
r_{23}^2 & = & R_{23}^2 + (v-u)^2 \nonumber
\end{eqnarray}

To average over $u$, we need the normalized probability
distribution $P(u|R_{12})$,  
\begin{eqnarray}
P(u|R_{12}) &=& \frac{1+ \xi_{gg}(\sqrt{u^2 + R_{12}^2})}{\int du~ 
(1+ \xi_{gg}(\sqrt{u^2 + R_{12}^2}))} \\
&=& \frac{1+ \xi_{gg}(\sqrt{u^2 + R_{12}^2})}{L_U + w_{12}(R_{12})}
\end{eqnarray}	
One can see that this distribution is improper in the sense that one cannot
normalize it over $(-\infty, \infty)$. Rather one needs to specify
a maximum radial separation between galaxies to do the integration, which 
we will call $L_U/2$. We will similarly take the maximum 
range for the $v$ integral as $L_V/2$.

Now we can write down the equation for the 2D mass density around galaxy pairs 
at separation $R_{12}$ by taking Equation \ref{eq:rho-pair-3D} and projecting 
this over $v$ and averaging over $u$,
\begin{eqnarray}
\Sigma(R_{13},R_{23}|R_{12})  = \rho_{crit} \Omega_m ~ \int\int dv~du~P(u|R_{12})~
\rho_m(r_{13},r_{23}|r_{12}) \nonumber \\
= \rho_{crit} \Omega_m ~ \int_{-L_V/2}^{L_V/2} dv~ \int_{-L_U/2}^{L_U/2} du P(u|R_{12}) 
\left[ 1 + ~ \frac{\xi_{gm}(r_{13}) +\xi_{gm}(r_{23}) 
+\zeta_{ggm}}{1 + \xi_{gg}(r_{12})} ~ \right]  \nonumber \\
= \rho_{crit} \Omega_m~  \left[ L_V + \frac{L_U}{L_U + w_{gg}(R_{12})}
(w_{gm}(R_{13}) + w_{gm}(R_{23})) + \frac{z_{ggm}}{L_U + 
w_{gg}(R_{12})} \right] \nonumber \\
\label{eq:sigma-with-L}
\end{eqnarray}
where we have defined the projected correlation functions,
\begin{eqnarray} 
w_{gm}(R_{13}) & = & \int dv~ \xi_{gm}(\sqrt{v^2+R_{13}^2}) \\
w_{gm}(R_{23}) & = & \int dv~ \xi_{gm}(\sqrt{v^2+R_{23}^2}) \\
w_{gg}(R_{12}) & = & \int du~ \xi_{gg}(\sqrt{u^2+R_{12}^2}) \\
z_{ggm}(R_{12},R_{23},R_{23}) & = & \int\int du~dv~ \zeta_{ggm}(u,v,R_{12},R_{23},R_{23})
\end{eqnarray}
From here on we will ignore the constant term $\propto L_V$ in Equation 
\ref{eq:sigma-with-L} since it is not observable with
shear measurements. Since this is the only term which involves $L_V$ 
explicitly, we can now let $L_V~\rightarrow~\infty $ 
so the integrals involving $v$ are now over all space. We can
rewrite Equation \ref{eq:sigma-with-L} without the constant as 
\begin{equation}
\Sigma(R_{13},R_{23}|R_{12})= \frac{\rho_{crit} \Omega_m}
{1+ \frac{w_{gg}(R_{12})}{L_U}}~\left[ w_{gm}(R_{13})+w_{gm}(R_{23}) 
+ \frac{z_{ggm}}{L_U}~\right]
\label{eq:sigma-simple}
\end{equation}
For lensing we need the dimensionless mass density $\kappa$, and so we
divide by $\Sigma_{crit}$. This Euclidean projection and division by
$\Sigma_{crit}$ implictly assumes that the correlation functions
do not extend over large enough distances to change the focusing strength.
A more general expression would include the radially varying 
$\Sigma_{crit}^{-1}$ inside the projection \citep{takada-jain:3pt-weak}. 
However, the correlation functions are only appreciable over several Mpcs, while 
$\Sigma_{crit}^{-1}$ typically varies over several hundred Mpcs
and so this is a reasonable approximation.

Given galaxy redshifts $z_1$ and $z_2$ and a 
source galaxy redshift $z_s$, Equation \ref{eq:sigma-simple} yields
\begin{equation}
\kappa(R_{13},R_{23}|R_{12})  = 
\frac{\rho_{crit} \Omega_m}{1+ \frac{w_{gg}(R_{12})}{L_U}}~
\left[ \frac{w_{gm}(R_{13})}{\Sigma_{crit}(z_1,z_s)} + 
\frac{w_{gm}(R_{23})}{\Sigma_{crit}(z_2,z_s)} 
+ \frac{z_{ggm}}{L_U ~\Sigma_{crit}(z_1,z_s)} \right] \label{eq:kappa-pair-1}
\end{equation}
Here we have assumed that $z_1 \simeq z_2$ in the third term, since
$z_{ggm}$ vanishes if this is not the case.
In most cases, we do not know the redshifts of the source galaxies precisely,
so we need to integrate over their probability distribution $P(z_s)$ 
which can usually be estimated, e.g., using photometric redshift estimates.
We define the univariate $\Sigma_{crit}(z_i)$ as
\begin{equation}
\Sigma_{crit}(z_i)^{-1} = \int dz_s P(z_s) \Sigma_{crit}(z_i,z_s)^{-1}
\end{equation}
Now we can rewrite Equation \ref{eq:kappa-pair-1} as
\begin{equation}
\kappa(R_{13},R_{23}|R_{12})  = 
\frac{\rho_{crit} \Omega_m}{1+ \frac{w_{gg}(R_{12})}{L_U}}~
\left[ \frac{w_{gm}(R_{13})}{\Sigma_{crit}(z_1)} + \frac{w_{gm}(R_{23})}
{\Sigma_{crit}(z_2)} 
+ \frac{z_{ggm}}{L_U ~\Sigma_{crit}(z_1)} \right] \label{eq:kappa-pair-2}
\end{equation}
This is the main expression we will use to predict the shear around pairs of
galaxies separated by $R_{12}$. N-body simulations will be used to 
calculate the projected galaxy-mass correlation functions, $w_{gm}$,
and $z_{ggm}$, and Equation \ref{eq:kappa-pair-2} then predicts 
$\kappa(R_{13},R_{23}|R_{12})$ around galaxy pairs for a survey
of a given depth. 

Equations \ref{eq:lens-potential} to \ref{eq:def-shear}
tell us how to take a convergence $\kappa$ and compute the resulting shear
map. Furthermore we can apply the KS algorithm to the shear map to reconstruct 
the $\kappa$ map. Equation \ref{eq:kappa-pair-2} also tells us how to interpret this reconstructed
$\kappa$ map in terms of projected correlation functions. Since the
term $w_{gg}(R_{12})$ is just the projection of the galaxy-galaxy 
correlation function $\xi_{gg}$, which we can measure, 
and since we can also will measure $w_{gm}$ by the 
usual two-point galaxy-galaxy lensing method (described in Section
\ref{section:gal-gal-lensing}) , 
we can measure the 2D three-point function $z_{ggm}$ directly from these
other measured functions and the reconstructed $\kappa$. 

If one lets the maximum radial galaxy separation, $L_U$, go to
infinity, one does not get any contribution from $z_{ggm}$. 
This just says that with an infinite background of galaxies,
the ratio of physical galaxy pairs to random projected
galaxy pairs will go to zero. If one's goal is to measure the
GGM3PCF, then one should not
make this measurement around pairs with widely different
redshifts: they only give contributions from their projected
two-point terms $w_{gm}(R_{13})$ and $w_{gm}(R_{23})$, which 
provides no new information beyond what one has already 
measured with two-point galaxy-galaxy lensing.
 
Instead one should set $L_U$ no larger than the scale of measurable galaxy
correlations. A good value might be $30 h^{-1}$ Mpc, since the
2PCFs are small beyond that and dominate
over the 3PCF though one might want to make $L_U$ larger
when making measurements at larger scales. At a minimum,
one will want $L_U$ greater than a few times the maximum
projected scales at which one is making the measurements.
In fact one could choose $L_U$ smaller to attempt to
maximize the signal-to-noise ratio of the measurement. However
at smaller scales redshift distortions make
the interpretation of radial distance problematic and so this 
probably will not prove useful. Making cuts on
redshift differences at any scale should really be interpreted
as a ``fuzzy selection'' in physical space. This selection function
$\phi(u)$ can be modeled with N-body simulations and incorporated
into the integrals above, rather than assume the top-hat form, 
to better interpret the measurements.

We can now estimate at which scales the three-point function $z_{ggm}$ 
will dominate over the two-point terms in Equation \ref{eq:kappa-pair-2}. 
For simplicity we choose the isosceles triangle configuration so that 
$R_{12} = R_{13} = R_{23} = R$. Let us model the three 2PCFs
as power laws, $\xi_{gg}(r) = \xi_{gm}(r) = (r/r_0)^{- \gamma}$, 
ignoring bias parameters which
are of order unity. We will take $r_0=6h^{-1}$ Mpc and $\gamma=1.8$.
The three projected 2PCFs are all then given by $w(R)= 3.7 r_0 (r_0/R)^{\gamma -1}$.
We will further assume that the projected three-point function $z_{ggm}$ can be
written as $z_{ggm}=Q [w_{gg}(R) w_{gm}(R) + w_{gg}(R) w_{gm}(R) + w_{gm}(R) w_{gm}(R)]$,
with $Q$ approximately constant. From Equation \ref{eq:sigma-simple}
we can write the three-point mass density $\Sigma(R)$ for this configuration as
\begin{equation}
\Sigma(R)=\frac{\rho_{crit} \Omega_m}{1+ \frac{3.7~r_0}{L_U} (\frac{r_0}{R})^{\gamma-1}} ~
3.7~r_0 \left( \frac{r_0}{R} \right) ^{\gamma-1} \left[ 2 + 3Q \frac{3.7~r_0}{L_U}
\left( \frac{r_0}{R} \right)^{\gamma-1}\right] \label{eq:isoc-approx}
\end{equation}
Since the usual two-point contribution from each galaxy,
which we will write here as $\Sigma_2(R)$, is 
$\Sigma_2(R)=\rho_{crit} \Omega_m~ 3.7~ r_0 (r_0/R)^{1-\gamma}$ 
(Equation \ref{eq:proj-sigma}), we can write Equation \ref{eq:isoc-approx} as 
\begin{equation}
\Sigma(R)=\frac{\Sigma_2(R)}{1+ \frac{3.7~r_0}{L_U} (\frac{r_0}{R})^{\gamma-1}} ~
\left[ 2 + 3Q \frac{3.7~r_0}{L_U} \left( \frac{r_0}{R} \right) ^{\gamma-1} \right] 
\end{equation}
The first term in brackets comes from the two-point terms 
in Equation \ref{eq:sigma-simple}. The next term is due to the three-point function
and equals the first term at $R=R_{eq}$, where
\begin{equation}
R_{eq}= r_0 \left(\frac{2 L_U}{3.7~3~Q r_0}\right)^{\frac{1}{\gamma-1}} \approx 0.9 h^{-1} \mbox{Mpc}
\end{equation}
where we have taken $Q=1$ and $L_U=30 h^{-1}$Mpc. For small scales $R \ll R_{eq}$,
where the three-point term dominates, we see that $\Sigma(R) \simeq 3~Q ~\Sigma_2(R)$,
independent of $L_U$. In general we want to subtract off the two-point term
so we define the reduced three-point surface mass density 
\[ \Sigma_3(R) = \Sigma(R) - \frac{2 \Sigma_2(R)}{1+ \frac{R_0}{L_U} (\frac{r_0}{R})^{\beta}} \]
We see that at $R=10 h^{-1}$ Mpc, $\Sigma_3(R) = 0.5 ~\Sigma_2(R)$
and at $R=100 h^{-1}$ Mpc, $\Sigma_3(R) = 0.1 ~\Sigma_2(R)$. 
Since we can measure the GM2PCF to at least
$R=10 h^{-1}$ Mpc in the SDSS \citep{sheldon:gmcf}, we should be able to 
measure the GGM3PCF if we have a similar number
of galaxy-pairs as we have galaxies. We will study this further in Section
\ref{section:sig-noise}.


\section{Halo occupation models}
\label{section:hod}
The halo occupation distribution models (HOD) provide a simplified
prescription for populating simulated or analytic
dark matter halos with galaxies. In the HOD 
framework \citep{berlind-weinberg:halo1,berlind:halo2}, 
one makes the assumption that the number of galaxies, $N$, that 
occupy a dark matter halo is drawn from a 
distribution $P(N|M)$ that only depends on $M$, the mass
of the halo. A priori, one might imagine that this is too restrictive and that the
probability distribution could depend on other parameters such as the shape of the
halo,  the formation time, the local density of other halos, or a host of other
parameters. However, hydrodynamical simulations indicate that halo mass 
is the controlling factor. 
Once one has determined the number of galaxies in each halo, one
needs to decide on how to distribute the galaxies within each halo. 
Together these two choices define a complete halo occupation prescription.

The chief usefulness of the HOD model for large scale structure studies is
that it provides a model for galaxy bias and therefore allows one to compute the various 
galaxy correlation functions in a straightforward manner.
This technique was first used with
randomly placed clusters with specified radial 
profiles by \cite{neyman-scott:halos}
to predict clustering statistics and also by \cite{mcclelland-silk:halos,peebles:halos} 
used similar models to compute
the galaxy correlation function. More modern treatments 
benefit from knowledge of the halo mass function 
\citep{press-schecter:halos,sheth-tormen:mass-function,
jenkins:mass-function,white:mass-function}
and halo mass profiles \citep*{nfw:profile} that are seen in N-body simulations. 
Another reason for the recent revival of this technique is the success of 
semi-analytic models of galaxy formation 
\citep{white-rees:sams,white-frenk:sams,lacey-cole:sams,
kauffmann-white:sams}. These models, which try to 
simulate the complex physics of galaxy formation, produce 
HODs which when combined with N-body dark matter simulations, result in 
a galaxy auto-correlation function that resembles the observed 
power law \citep{benson:sams,kauffmann:sams}. Due to this
improved knowledge of the mass function, 
halos profiles, and the HOD, there has been much recent work 
on computing the various N-point galaxy correlation functions 
with the halo model formalism
\citep{seljak:halos,ma-fry:halos,ma-fry:2pt3pt,peacock-smith:halos,scoccimarro:howmany,
takada-jain:3pt,scranton:halos}.

In the purely analytic halo model formalism, one combines the HOD
with an analytic halo mass function and analytic halos profiles that are fit from 
N-body simulations. Another approach is to take an N-body
dark matter simulation, identify halos, and populate these halos directly
with galaxies using the HOD prescription. This is the method we will use in
this study. This method yields a set of 
dark matter positions and a set of galaxy positions
that one can use to calculate the various correlation and cross
correlation functions. It has the advantage 
that it samples a realistic distribution
of halo shapes and structures not captured by the analytic method,
which usually assumes spherical halos with a universal profile.
In particular, while the constraint of spherical halos should not 
matter for the 2PCFs, it may be more important for the higher
N-point functions since they are anisotropic. In fact, the
analytic halo models do not agree precisely with N-body
predictions for the 3PCF, because the former are not hierarchical
on small scales.

However, populating dark matter simulations with galaxies also has its drawbacks.
One problem is that if one wants to change the dark matter power spectrum or
vary other cosmological parameters, one needs to run another N-body
simulation. With the purely analytic technique, one can calculate the various
correlation functions simply with numerical integrations; to change
cosmological parameters one just has to re-run these integration routines,
which take far less CPU time. Thus with the purely analytic technique one
can explore parameter space more efficiently. Also with the purely analytic
technique one can compute the various statistics with no errors. With the
N-body population technique one always has shot noise and sample variance.
However if one has a large enough volume and enough particles and galaxies,
one can make predictions that will have theoretical errors smaller than errors coming
from real survey data. Another thing to note about the analytic models is that
the 2PCFs are easily computed with simple one dimensional
numerical integrations but the higher order N-point functions 
can require much higher order numerical integration. 
For example, to compute the full three-point function 
one must compute a seven-dimensional numerical integral and since this is 
generally intractable, one has to resort to some approximation methods
\citep{takada-jain:3pt}. In addition to the extra numerical computation, 
the mathematical expressions become significantly more complicated and require 
more assumptions about parameters and input from simulations.

\section{The N-body simulation}

The N-body dark matter
simulation used for this study is the public simulation made available at
http://pac1.berkeley.edu/Sim1/. This simulation has $512^3$ particles in a 
periodic cube of length 300 $h^{-1}$ Mpc. The simulation is evolved from $z=60$ to $z=0$
using a TreePM code \citep{white:mass-function,white:planck}. 
The initial conditions are set using the Zel'dovich
approximation to displace the particles from the uniform grid.
The cosmological parameters used are $\Omega_M=0.3$,
$\Omega_{\Lambda}=0.7$, $h=0.7$, $n=1$, $\Omega_b~h^2 =0.02$ and $\sigma_8=1$.
The CDM transfer function used is that of \cite{eisenstein-hu:transfer} with the 
baryon wiggles smoothed over. The simulation has an effective Plummer
force-softening scale of 20 $h^{-1}$ kpc which is fixed in comoving coordinates.
The mass of each dark matter particle is $1.7 \times 10^{10} h^{-1} M_{\sun}$. The simulation
outputs three positions and three velocities for each particle 
but only the positions are used for this
study. The dark matter halos are found using a Friends-of-Friends (FoF) algorithm
\citep{davis:fof} with a linking length of 0.2 in units of the mean inter-particle
separation. 

The N-body simulation is then populated with galaxies at $z=0$
using halo occupation distribution (HOD) techniques based on 
\cite{berlind-weinberg:halo1}, courtesy of Berlind. 
The HOD parameters are chosen to produce a galaxy 2PCF
that matches $\xi_{gg}(r)$ of the SDSS
volume-limited sample of \cite{zehavi:departure}
for absolute magnitudes $M_r < -21$. There are 
123081 galaxies corresponding to a comoving number density
of $n_g = 4.56 \times 10^{-3} h^3 /Mpc^{3}$.
This number density corresponds to an 
absolute magnitude cut $M_r < -20.2$ according
to the luminosity function of \cite{blanton:luminosity}.
The $P(N|M)$ used is one that has a mean 
$\langle N(M) \rangle$ which is as follows
\begin{equation}
 \langle N(M) \rangle = \left\{
\begin{array}{ll}
0 & M < M_{min} \\
1 & M_{min} < M < M_1 \label{eq:hod-mass} \\ 
(M/M_1)^{\beta} & M > M_1
\end{array} 
\right.
\end{equation}
The parameters we use are $M_{min}=1.09 \times 10^{12} h^{-1} M_{\sun}$ and 
$M_1=9.97 \times 10^{12} h^{-1} M_{\sun}$, with $\beta = 0.83$.

Equation \ref{eq:hod-mass} gives the mean number of galaxies for a halo of mass $M$. 
We must allow some scatter when we choose the number
of galaxies for each halo. Since we must choose an integer number, 
we choose either one of the integers bracketing the mean with
relative probabilities that keep the mean constrained to equal
Equation \ref{eq:hod-mass} when the mean is non-integer. When 
$\langle N \rangle$ is an integer, we simply choose that 
number with no variance. This leads to a variance 
that is substantially sub-Poisson and in fact is
the minimum variance that is not identically zero. To be more specific,
if the mean galaxy number is $\overline{N} = N + \epsilon$ with $N$ an integer
and $\epsilon$ the fractional part, then we choose $N$ with probability 
$1-\epsilon$ and $N+1$ with probability $\epsilon$.
One can easily show that the resulting variance is 
$\sigma^2=\epsilon( 1 - \epsilon )$. The variance therefore depends on where between
two integers $\overline{N}$ happens to lie, and it is maximized 
at $\sigma^2=1/4$, half way between two integers ($\epsilon=0.5$).

Once the number of galaxies is chosen for a given halo, the
first galaxy is always placed at the halo center of mass 
and the remaining galaxies are each placed randomly 
at the positions of one of the other dark matter particles
in that halo. For more details of this 
HOD model see \cite{zehavi:departure,berlind-weinberg:halo1}.

\section{Correlation functions from the simulation}
\subsection{The 3D two-point functions}

We measure the 3D two-point correlation functions in the simulation using
two different methods. On large scales we bin the data onto
a cube of $256^3$ pixels, Fourier transform to measure
the power spectrum, and then transform back to get the 
correlation function. This method works well
but because of the finite resolution imposed by memory limitations,
it is limited to measuring only on large scales. This method makes
reliable measurements of $\xi(r)$
on scales 2$h^{-1}$ Mpc to 50$h^{-1}$ Mpc.

For smaller scales, we do simple pair counting to estimate $\xi(r)$.
We break the entire region
of data into $N_{seg}^3$ different segments and write each to a different file.
This enables one to search quickly for pairs since one only needs to read in
the neighboring 27 (or 9 in 2D) segments to measure pairs within a separation
of 1 segment. We typically use $N_{seg}=100$, which gives a maximum 
search radius of 300/100=3$h^{-1}$ Mpc. This range overlaps the range
of the Fourier method, and we find excellent agreement between the
two methods in the overlap region.

\begin{figure}
\begin{center}
\epsfxsize=12.0cm \hspace{-2cm} \epsfbox{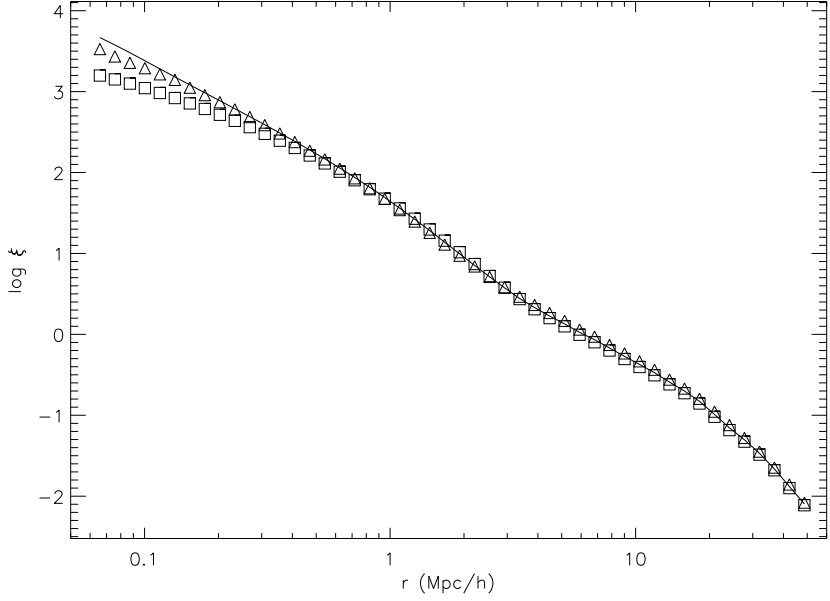}
\end{center}
\caption[The 3D two-point correlation functions] {
The three 3D two-point correlations functions
from 0.06$h^{-1}$Mpc to 50 $h^{-1}$Mpc in the N-body
simulation. The solid curve denotes
$\xi_{gm}(r)$, the triangles denote $\xi_{gg}(r)$, and the 
squares denote $\xi_{mm}(r)$.
\label{fig:corr-func-all-2pt-3d}}
\end{figure}

The 3D correlation functions are shown in Figure \ref{fig:corr-func-all-2pt-3d}.
These plots show that on scales greater than about 1$h^{-1}$ Mpc,
the three curves trace
each other fairly well. However, on smaller scales both $\xi_{gm}(r)$
and $\xi_{gg}(r)$ are biased positively with respect to the mass. The 
bias of $\xi_{gg}(r)$ is expected, since it is neccesary
to have substantial bias at small scales
to fit the observed $\xi_{gg}(r)$, which is known to be close to a power law.
Also one can see that on these small scales, $\xi_{gm}(r)$ is greater than 
$\xi_{gg}(r)$. Both of these qualitative results are expected with the
HOD framework. On small scales, the 2PCFs are dominated 
by the contribution of galaxies in the smaller halos 
\citep{seljak:halos}, which are the most abundant; 
many of them contain only one galaxy at the center of the halo. 
Because the galaxy is at the center,
$\xi_{gm}(r)$ traces the cusp of the dark matter halo and will therefore 
be steepest at small scales. $\xi_{mm}(r)$ is smoothed out somewhat, because 
this is basically the halo profile convolved with itself. In the case of 
$\xi_{gg}(r)$, it is suppressed at small scales due to the fact that small
halos have a $P(N|M)$ that excludes a second galaxy, so $\xi_{gg}$ only gets
contributions from bigger halos and the two-halo term (galaxies in different
halos). This argument explains qualitatively why $\xi_{gm}(r)$ is the largest at
small scales. The most likely reason why $\xi_{gg}(r)$
is larger than $\xi_{mm}(r)$ is that the $P(N|M)$ gives
more weight to smaller halos for $\xi_{gg}(r)$ and one
galaxy is always at the center. These effects must be more important
than the exclusion effect described above, for $\xi_{gg}(r)$ to be larger than
$\xi_{mm}(r)$. 

\subsection{Scale dependent bias}

It is conventional to define the bias parameters as ratios of the
correlation functions or power spectra. The bias parameters
can therefore be scale dependent; in fact scale dependent
bias is necessary if one is to get a nearly power-law galaxy correlation
function (that one sees in galaxy surveys), from a non-power-law mass 
correlation function that one sees in N-body simulations.
We define the galaxy bias parameter as $b^2(r)=\xi_{gg}(r)/\xi_{mm}(r)$
and the galaxy-mass cross correlation coefficient 
$\mbox{r}(r) = \xi_{gm}(r)/\sqrt{\xi_{mm}(r) \xi_{gg}(r)}$ \citep{pen:bias}.
From these definitions one can also see that 
$b(r)/\mbox{r}(r)= \xi_{gg}(r)/\xi_{gm}(r)$. The
ratio $b/\mbox{r}$ is directly measurable from combining 
galaxy-galaxy lensing measurements with 
galaxy auto-correlation measurements \citep{sheldon:gmcf}.

\begin{figure}
\begin{center}
\epsfxsize=12.0cm \hspace{-2cm} \epsfbox{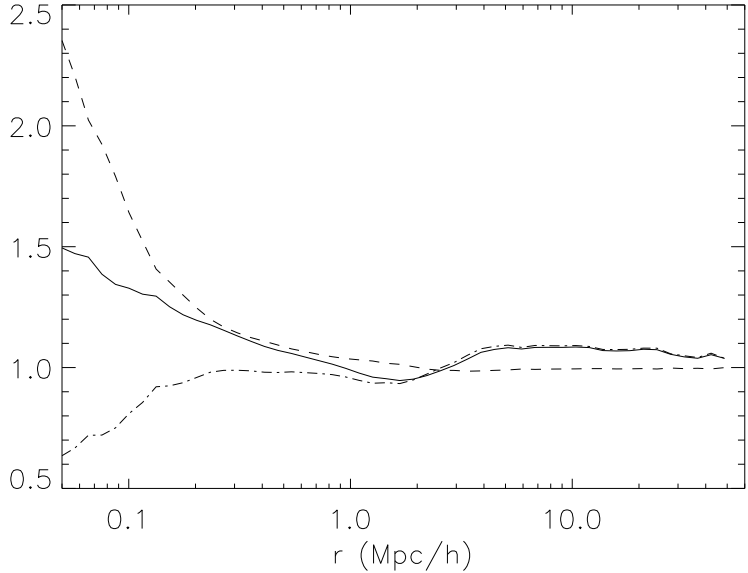}
\end{center}
\caption[The bias parameters] {
The bias parameters as a function of scale. The solid curve
is $b=\sqrt{\xi_{gg}/\xi_{mm}}$. The dashed curve is $\mbox{r}=\xi_{gm}/\sqrt{\xi_{gg} \xi_{mm}}$
and the dot-dashed curve is $b/\mbox{r}$.
\label{fig:bias-r-b-bor}}
\end{figure}

These three quantities $b$, $\mbox{r}$, $b/\mbox{r}$ for the simulation are
shown in Figure \ref{fig:bias-r-b-bor}. The bias parameter
$b$ starts at about 1.5 at $r=0.05 h^{-1}$ Mpc, decreases to about 0.95 at
about $r=1.5 h^{-1}$ Mpc and then increases slightly at larger separation.
This behavior is most likely due to
the transition from the one-halo term to the two-halo term. At large scales,
$b$ plateaus at about 1.05, though there is 
some indication that it falls off slowly toward larger scales.
The $\mbox{r}$ parameter is decreasing monotonically with $r$ out to 
$2~h^{-1}$ Mpc and beyond that it is flat at $\mbox{r}=1$.

The ratio $b/\mbox{r}$ is shown as the dot-dashed curve. It is 
fairly flat from $r=0.2~h^{-1}$ Mpc to
$1.0~h^{-1}$ Mpc and then rises to a slightly higher plateau
by $3.0 h^{-1} Mpc$; beyond $r=2~h^{-1}$ Mpc $b/\mbox{r}=b$ since
$\mbox{r}=1$.

On small scales $r < 0.2 h^{-1}$ Mpc, $b/\mbox{r}$
drops by almost half. This scale dependence of
$b/\mbox{r}$ would not yet be apparent in SDSS lensing data but it should become 
measurable in future surveys. The overall amplitude 
$b/\mbox{r} \approx 1$ agrees with SDSS galaxy-galaxy lensing 
measurements \citep{sheldon:gmcf} for $\Omega_m \approx 0.3$.

\subsection{The 2D two-point functions}

\begin{figure}
\begin{center}
\epsfxsize=12.0cm \hspace{-2cm} \epsfbox{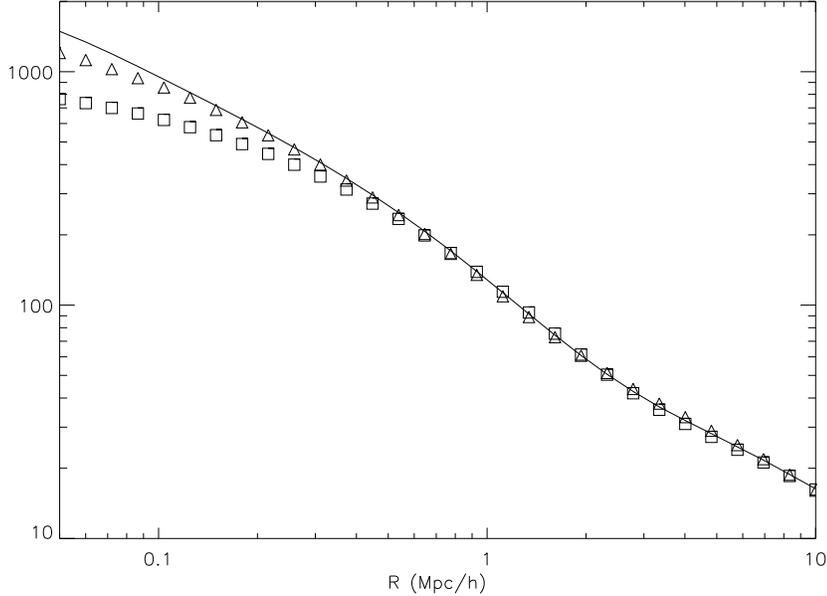}
\end{center}
\caption[The projected two-point correlations functions] {
The three projected two-point correlations functions
from $R=0.05~h^{-1}$Mpc to 10 $h^{-1}$Mpc. The solid curve denotes
$w_{gm}(R)$, triangles denote $w_{gg}(R)$, and the 
squares denote $w_{mm}(R)$.
\label{fig:corr-func-all-2pt-2d-proj}}
\end{figure}

We also measure the 2D two-point functions where we first project the
data (i.e. ignore the third coodinate) and measure correlation functions
of the 2D densities. However these are simply projected
versions of the 3D two-point functions and the analytic 
projections of the 3D functions
agree very well with the measured 2D functions. We note that projected
correlation functions, which are usually referred to as $w_p$, have dimensions
of length. If one measures a 2D correlation function by first projecting the points
to 2D and then computing the 2D correlation function $w_{2D}$ in the usual way, 
the result is dimensionless and it is equal to $w_p$/L, where L is the dimension 
of the box in the radial direction. When we refer simply to $w$, we mean $w_p$,
which has dimensions of length. Figure \ref{fig:corr-func-all-2pt-2d-proj} 
shows the three projected correlation functions.

\subsection{The 3D three-point functions}

Next we measure the 3D 3PCFs $\zeta_{ggg}$ and $\zeta_{ggm}$.
Our main purpose for measuring these functions is simply to 
check our results against other published results of $\zeta_{ggg}$. 
For the most part we will be interested in the 2D three-point 
functions, since they are directly measurable with lensing.
The three-point function depends on the three triangle sides,
however it is usually parametrized with the three variables being
two side lengths $r$ and $q$ and the angle $\psi$ at the intersection of
these two sides. Motivated by the hierarchical form of Equation \ref{eq:zggg},
it is customary to divide $\zeta_{ggm}$ by the usual
symmetric combinations of 2PCFs
to define the scaled 3PCF $Q_{ggm}$ for two galaxies and a mass particle 
\begin{equation}
Q_{ggm} \equiv \frac{\zeta_{ggm}}{\xi_{gg}(r_{12}) \xi_{gm}(r_{13}) 
+ \xi_{gg}(r_{12}) \xi_{gm}(r_{23}) 
+ \xi_{gm}(r_{13}) \xi_{gm}(r_{23})}
\label{eq:hier-3d}
\end{equation}
and likewise for three galaxies
\begin{equation}
Q_{ggg} \equiv \frac{\zeta_{ggg}}{\xi_{gg}(r_{12}) \xi_{gg}(r_{13}) 
+ \xi_{gg}(r_{12}) \xi_{gg}(r_{23}) 
+ \xi_{gg}(r_{13}) \xi_{gg}(r_{23})}
\label{eq:hier-3d-gals}
\end{equation}

We compute the 3D 3PCFs only on small scales by first creating a 
galaxy pair catalog; for each pair, we search for neighboring
galaxies or dark matter particles to complete the triangle.
We limit the first galaxy pair separation $r$ from 0.1 $h^{-1}$ Mpc to
0.9 $h^{-1}$ Mpc and search around the center of the pair to a maximum 
radius of 3.0 $h^{-1}$ Mpc.

\begin{figure}
\begin{center}
\epsfxsize=12.0cm \hspace{-2cm} \epsfbox{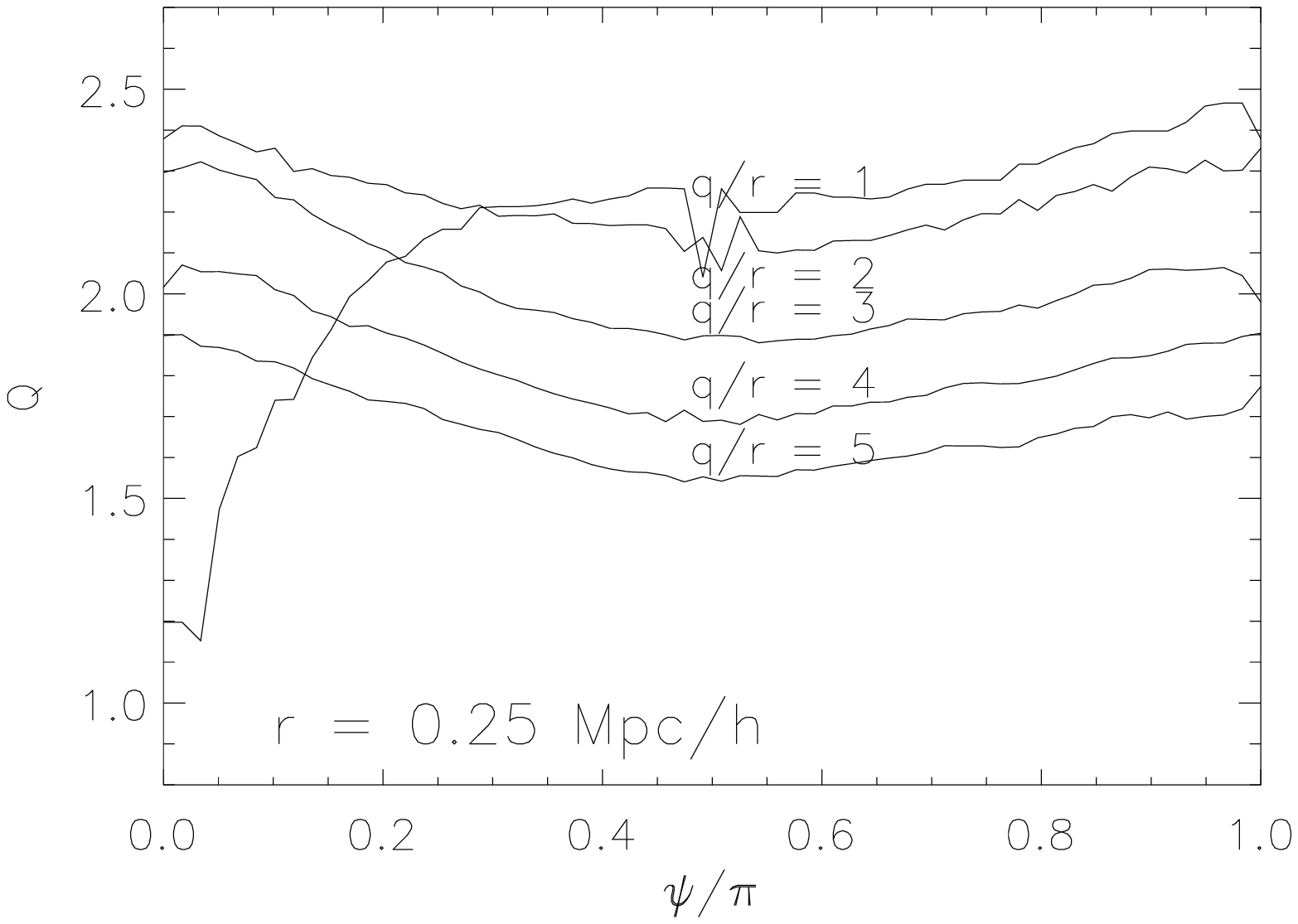}
\end{center}
\caption[The 3D $Q_{ggm}$ for $0.25 h^{-1}$ Mpc] {
The 3D $Q_{ggm}$ as a function of angle $\psi$
for galaxy separation $r_{12} = 0.25 h^{-1}$ Mpc and for
various $q/r$ ratios. 
\label{fig:Q-theta-sep-0.25}}
\end{figure}

\begin{figure}
\begin{center}
\epsfxsize=12.0cm \hspace{-2cm} \epsfbox{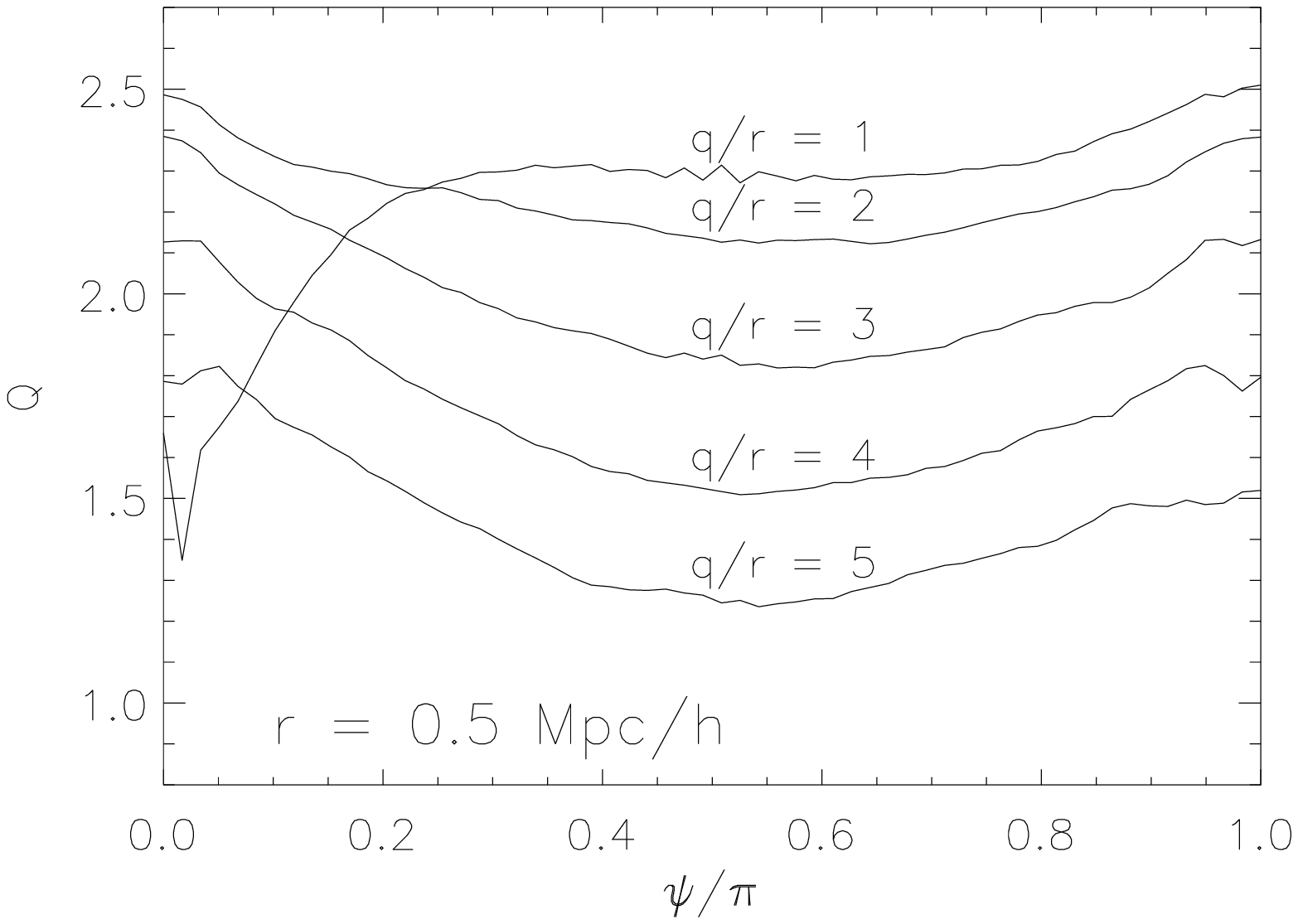}
\end{center}
\caption[The 3D $Q_{ggm}$ for $0.5 h^{-1}$ Mpc] {
The 3D $Q_{ggm}$ as a function of angle $\psi$
with galaxy separations $r_{12}=0.5 h^{-1}$ Mpc and for
various $q/r$ ratios. 
\label{fig:Q-theta-sep-0.5}}
\end{figure}

\begin{figure}
\begin{center}
\epsfxsize=12.0cm \hspace{-2cm} \epsfbox{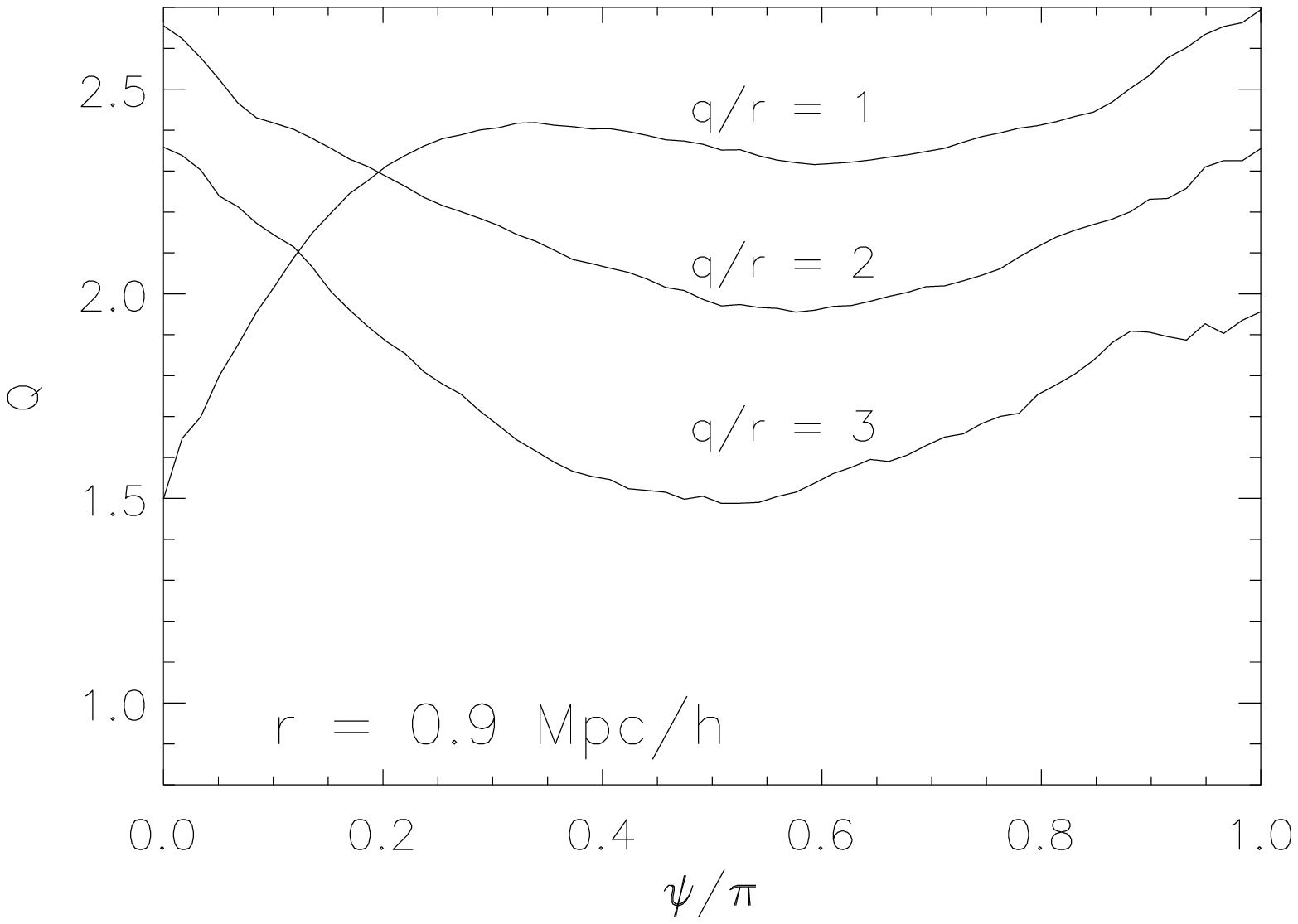}
\end{center}
\caption[The 3D $Q_{ggm}$ for $0.9 h^{-1}$ Mpc] {
The 3D $Q_{ggm}$ as a function of angle $\psi$
with galaxy separations $r_{12}=0.9 h^{-1}$ Mpc and for
various $q/r$ ratios. 
\label{fig:Q-theta-sep-0.9}}
\end{figure}

Figure \ref{fig:Q-theta-sep-0.25} shows the reduced function $Q_{ggm}$
for galaxies with separations $r = 0.25 \pm 0.04 h^{-1}$ Mpc for 5
side ratios $q/r$ from 1 to 5. Figure \ref{fig:Q-theta-sep-0.5}
is the same plot but for galaxies with separations $r  = 
0.5 \pm 0.04 h^{-1}$ Mpc, and Figure 
\ref{fig:Q-theta-sep-0.9} for galaxies with 
separations $r = 0.9 \pm 0.04 h^{-1}$ Mpc.

\begin{figure}
\begin{center}
\epsfxsize=12.0cm \hspace{-2cm} \epsfbox{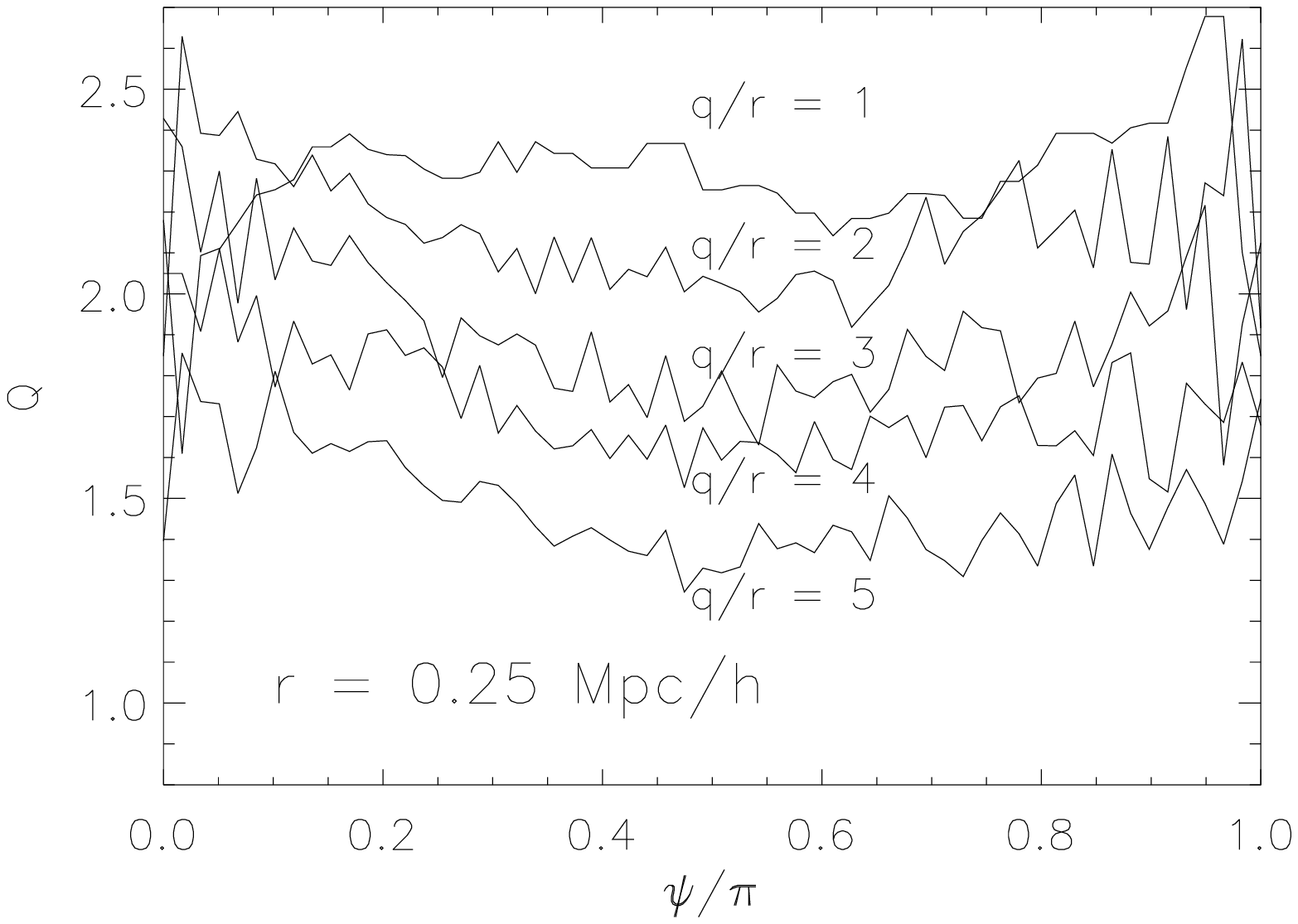}
\end{center}
\caption[The 3D $Q_{ggg}$ for $0.25 h^{-1}$ Mpc] {
The 3D $Q_{ggg}$ as a function of angle $\psi$
with galaxy separations $r_{12}=0.25 h^{-1}$ Mpc and for
various $q/r$ ratios. 
\label{fig:Q-theta-sep-0.25-gals}}
\end{figure}

\begin{figure}
\begin{center}
\epsfxsize=12.0cm \hspace{-2cm} \epsfbox{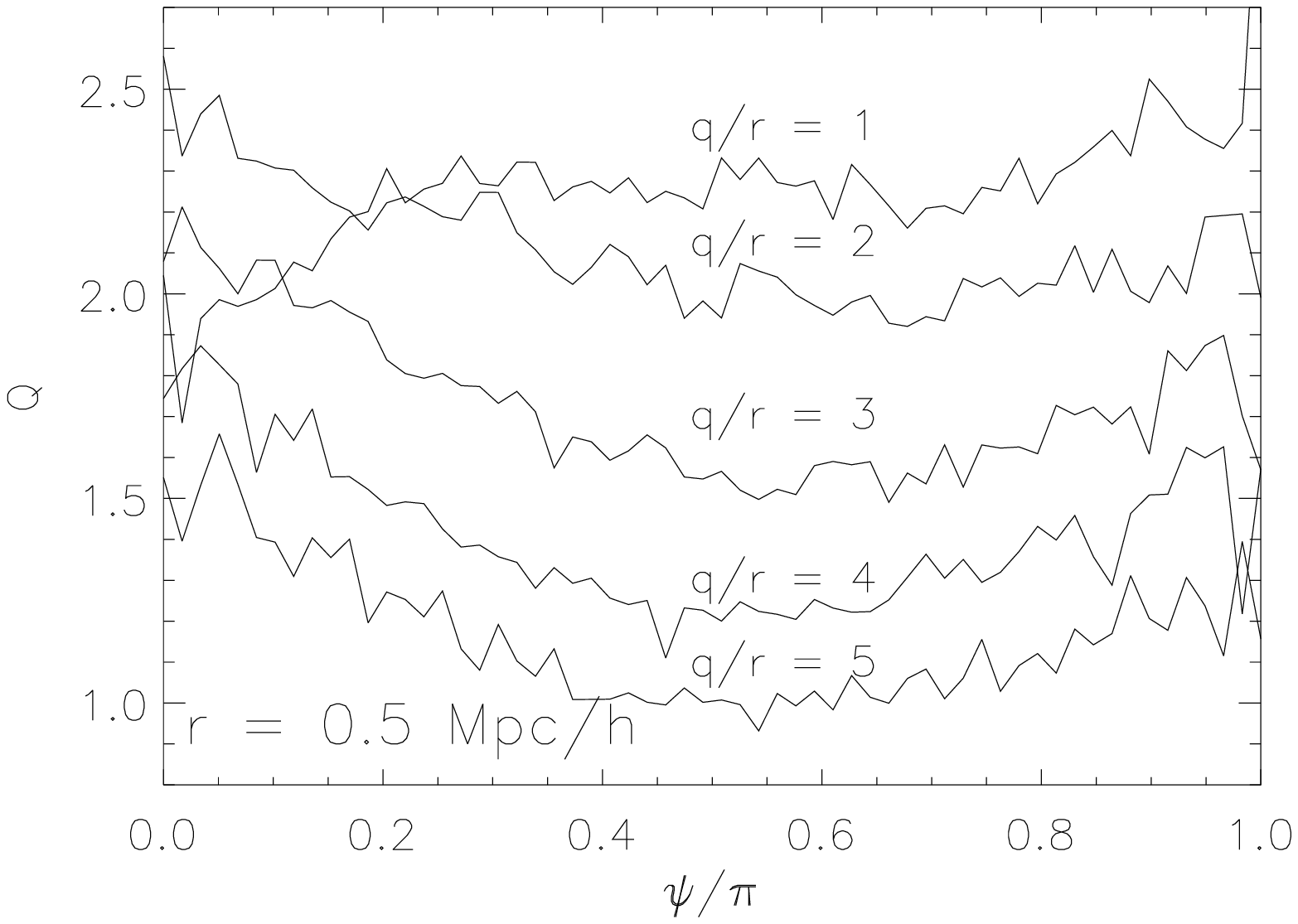}
\end{center}
\caption[The 3D $Q_{ggg}$ for $0.5 h^{-1}$ Mpc] {
The 3D $Q_{ggg}$ as a function of angle $\psi$
with galaxy separations $r_{12}=0.5 h^{-1}$ Mpc and for
various $q/r$ ratios. 
\label{fig:Q-theta-sep-0.5-gals}}
\end{figure}

\begin{figure}
\begin{center}
\epsfxsize=12.0cm \hspace{-2cm} \epsfbox{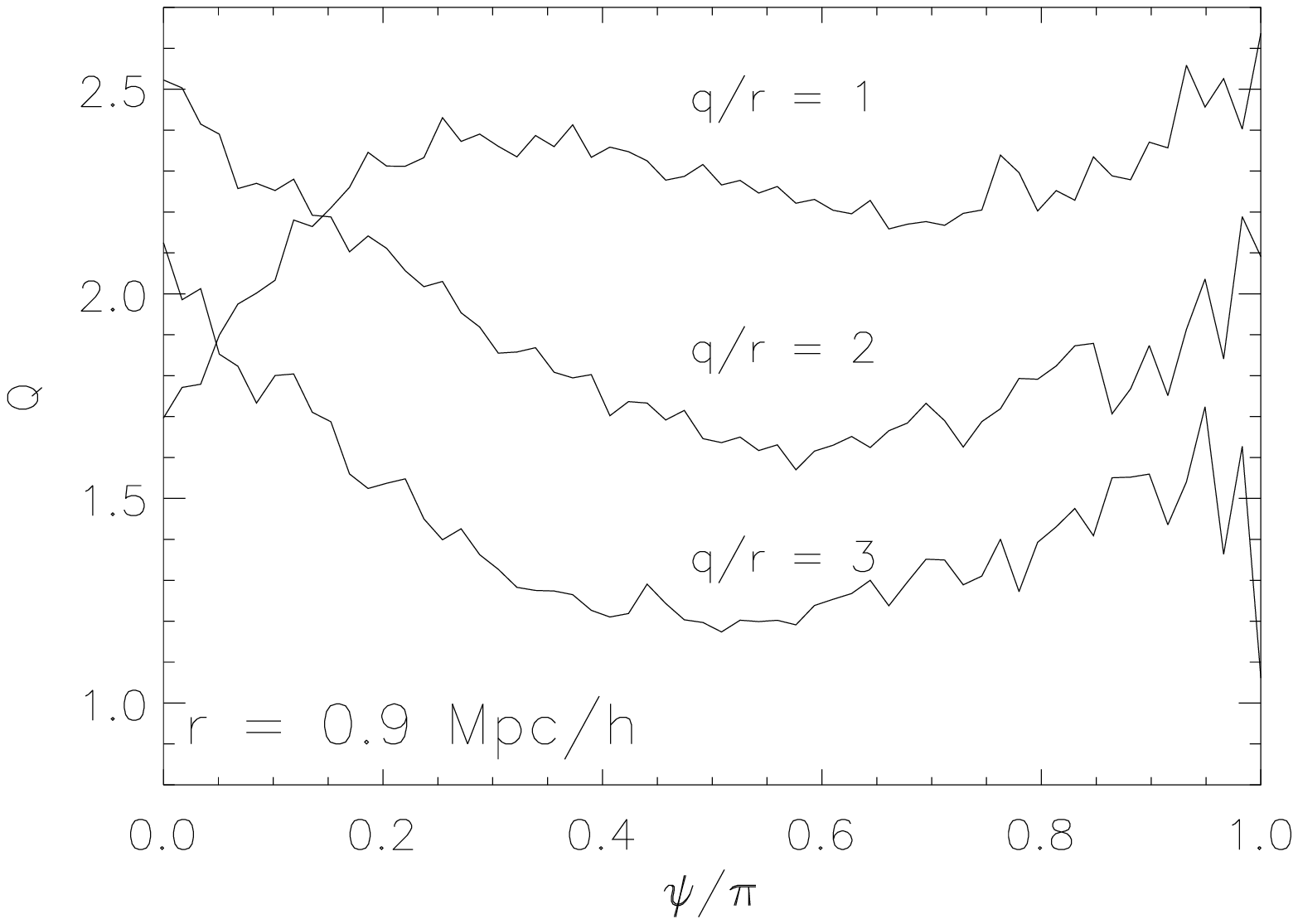}
\end{center}
\caption[The 3D $Q_{ggg}$ for $0.9 h^{-1}$ Mpc] {
The 3D $Q_{ggg}$ as a function of angle $\psi$
with galaxy separations $r=0.9 h^{-1}$ Mpc and for
various $q/r$ ratios. 
\label{fig:Q-theta-sep-0.9-gals}}
\end{figure}

We also plot the function $Q_{ggg}$, although this is much
noisier since there are many fewer galaxies than dark matter particles.
These three plots are Figures \ref{fig:Q-theta-sep-0.25-gals},~
\ref{fig:Q-theta-sep-0.5-gals}, and \ref{fig:Q-theta-sep-0.9-gals} 
for galaxy separations 0.25, 0.5, and 0.9 $h^{-1}$ Mpc respectively.

Although there is a lot of variation in these plots, there is a
generic trend that describes all of them. First, the hierarchical
scaling works fairly well, in that both $Q_{ggm}$ and $Q_{ggg}$
only vary between 1 and 3 over all of the plotted ranges. Except for the case
$q/r=1$, the amplitude of $Q$ is higher on either end when $\psi=0$ or
$\psi=\pi$ and lower in the middle when $\psi=\pi/2$ and is fairly symmetric
about $\psi=\pi/2$. This indicates a
small preference for collinear triangles. This preference for collinear
structures is predicted in perturbation theory 
and observed in galaxy surveys for much 
larger scales but with a much larger variation with $\psi$ 
\citep{fry:3pcf-hierarchy,takada-jain:3pt,feldman:bispectrum}.
Halo model calculations by \cite{takada-jain:3pt} (hereafter TJ) 
reveal a flatter curve with $\psi$ for 
their $Q_{ggg}$ then ours at $r=0.1 h^{-1}$ Mpc.
This small difference may be an indication of halo ellipticity, since
TJ are constrained to use spherical halos in their analytic calculation, or
it may simply be due to TJ using different parameters from the ones in this
simulation. The larger scale $r=1 h^{-1}$ Mpc plots of TJ shows more complicated
behavior where the curves for different $q/r$ are not self similar. 
This is due to the fact that all three terms , one-halo, two-halo, and three-halo,
are contributing at this scale and each term contributes differently for each
$q/r$. Our curves at $r=0.9 h^{-1}$ Mpc look 
different from theirs and not too different from our own at smaller scales.
This may indicate that in our simulations the one-halo term still dominates
at this scale whereas in TJ the two and three halo terms are becoming
important.

We also see that the $q/r > 1$ curves are fairly self-similar and simply decrease in
amplitude as $q/r$ increases. This is also seen in TJ at small scales. For the case
$q/r = 1$, we have somewhat different behavior: the curve is not symmetric about
$\psi=\pi/2$.
For large angles ($\psi \sim \pi$) the curves looks similar to the other
$q/r > 1$ curves but at $\psi$ close to zero Q decreases substantially.
These $\psi \sim 0$ triangles are such that the third galaxy or dark matter particle
is right on top of the first galaxy. So this decrease in Q indicates that the
hierarchical scaling does not work as well very close to the two galaxies. The value
of Q can get as small as 0.1 for cases where the third particle is very close to the
first (or second) galaxy, but it never goes negative at the scales we consider.
It is possible in principle for Q to be negative; for example, in TJ and also in perturbation theory 
\citep{frieman:3pt-APM}, Q can go as low as -4 on large scales ( $> 10 h^{-1}$ Mpc).
This decrease in Q at $\psi \sim 0$ is also seen in TJ and seems to be characteristic
of the one-halo term and probably is affected by the inner halo slope and the number
of small mass halos.

\subsection{The 2D three-point functions}

The 2D GGM3PCF is the object that
can be measured directly with weak lensing. As we described in 
Section \ref{section:measuring-higher} the 2D mass over density, 
$\Sigma(x,y)$, around pairs of galaxies separated by a fixed 
projected distance $R_{12}$ is given by Equation \ref{eq:kappa-pair-2},
which we rewrite as 
\begin{equation}
\Sigma(x,y) = \frac{\rho_{crit}~\Omega_m}{1+\frac{w_{gg}(R_{12})}{L_U}}
\left[~w_{gm}(R_{13})+w_{gm}(R_{23})+\frac{z_{ggm}(x,y)}{L_U} ~\right] 
\end{equation}
where 
\begin{eqnarray}
R_{13}(x,y) & = & \sqrt{(x+R_{12}/2)^2+y^2} \label{eq:R13-xy} \\
R_{23}(x,y) & = & \sqrt{(x-R_{12}/2)^2+y^2} \label{eq:R23-xy}
\end{eqnarray}
Here we have simply 
used the parameter $R_{12}$ for the galaxy separation and $x$, $y$ for
the Cartesian coordinates centered on the midpoint 
of the galaxy pair and rotated to make
the galaxy pair lie along the x-axis. These three parameters 
$R_{12}$, $x$, and $y$ specify a triangle connecting the 
two galaxies and a mass particle.
 
\begin{figure}
\begin{center}
\epsfxsize=12.0cm \hspace{-2cm} \epsfbox{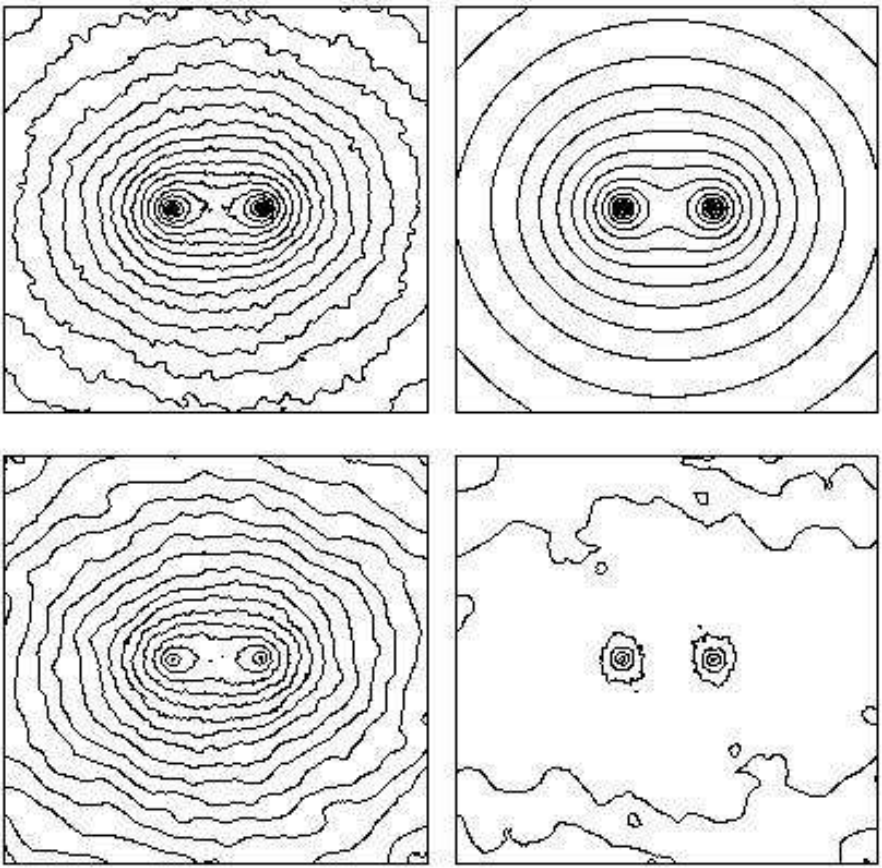}
\end{center}
\caption[The 2D mass density around pairs of galaxies separated by 
$R_{12}=0.9~h^{-1}$ Mpc] {
Top left: the 2D projected mass density
$\Sigma(x,y)$ for fixed projected galaxy separation
$R_{12}=0.9 h^{-1}$ Mpc. At small scales 
there is a bimodal distribution with
peaks around the locations of the two galaxies. At larger scales
the density has elliptical contours. Top right:
the contribution from the two point functions $w_{13} + w_{23}$.
These are simply computed from tabulated values for these functions
and so this has no apparent noise. Bottom left:
the galaxy-galaxy-mass three point function $z_{ggm}$. This
is obtained from taking $\Sigma(x,y)$ and subtracting off the constant 
and the two-point contributions (see Equation \ref{eq:sigma-with-L}).
Lower right: the reduced projected three point
function $q_{ggm}$ which is obtained by dividing $z_{ggm}$ by the
symmetrized combination of products of two-point functions
(Equation \ref{eq:q-2d-def}).
At large scales $q_{ggm}$ drops off slowly and retains
some ellipticity. At smaller scales it reveals holes around
the galaxies where $q_{ggm}$ drops off sharply. These features of
the halo models might not be present in reality since these models
do not include sub halos of dark matter or even baryonic mass
around the galaxy centers. All of these densities are log-scaled 
except $q_{ggm}$ and the length of these boxes are $4.24 ~h^{-1}$ Mpc on 
a side.  
\label{fig:sigma4-ggm-sep-0.9-side-4.24}}
\end{figure}

We display contour plots for the 2D density around pairs with 
$R_{12}=0.9 h^{-1}$ Mpc in Figure \ref{fig:sigma4-ggm-sep-0.9-side-4.24}. 
These boxes are 4.24 $h^{-1}$ Mpc on a side. This is a projection through the 
whole volume for all projected galaxy pairs, so here $L_U=300 h^{-1}$ Mpc.
The top left contour plot shows $\Sigma(x,y)$. The top right plot shows the
contribution coming from the 
two-point functions $w_{gm}(R_{13}(x,y))+w_{gm}(R_{23}(x,y))$.
This must be subtracted from $\Sigma(x,y)$ to get the reduced three-point
function. One can see how these terms are each radially symmetric about
the two galaxies, so their sum creates most of the bimodal pattern seen
in the first plot. The plot in the lower left is $z_{ggm}(x,y)$.
Even after subtracting off the $w_{13}+w_{23}$ parts, there is still
a bimodal pattern which is expected from the hierarchical model. Finally
the plot in the lower right shows the scaled three-point function,
$q_{ggm}(x,y)$, which is defined
\begin{equation}
q_{ggm}(x,y)=\frac{z_{ggm}(x,y)}{w_{gg}(R_{12}) ~w_{gm}(R_{13}) + w_{gg}(R_{12}) 
~w_{gm}(R_{23}) + w_{gm}(R_{13}) ~w_{gm}(R_{23})} \label{eq:q-2d-def}
\end{equation}
Here, the distances $R_{13}$ and $R_{23}$ are functions of $x$ and $y$
(Equations \ref{eq:R13-xy} and \ref{eq:R23-xy}) but $R_{12}$ is fixed.
At larger scales $q_{ggm}$, is fairly constant at 1.5 
although it drops off slowly. On small scales around each galaxy, 
there are holes where $q_{ggm}$ drops from $\sim 1.8$ on 
the outside to $\sim 0.6$ at the center.
This is similar to what we saw in 3D: the hierarchical scaling does not
appear to work well very close to each galaxy. This may be a feature
of the HOD models, since they do not associate sub-halos to each galaxy:
non-central galaxies do not have a mass excess around them
at small scales as do the central galaxies. In reality, it is probably
the case that non-central galaxies have small sub-halos of dark matter
around them. We do know at the very least that they have baryonic mass
tightly clustered around them in the form of stars, so the HOD models
without sub-halos must give incorrect galaxy-mass correlation function
at these small scales (a few kpc) where baryons dominate. 

\begin{figure}
\begin{center}
\epsfxsize=12.0cm \hspace{-2cm} \epsfbox{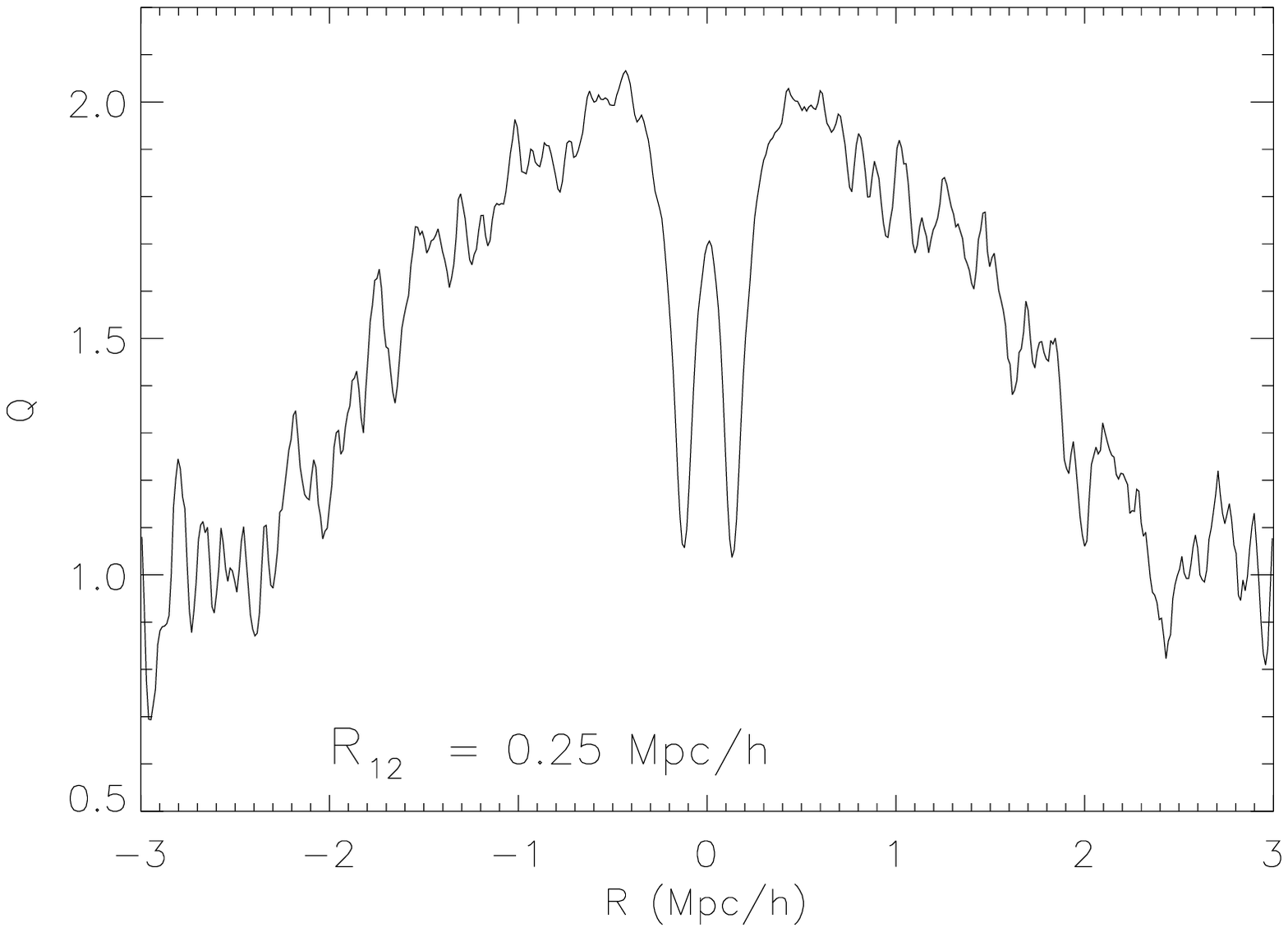}
\end{center}
\caption[A slice through the 2D $q_{ggm}$ for pairs with $R_{12}= 0.25~h^{-1}$ Mpc] {
A slice through the 2D $q_{ggm}(x,y)$ along the $y=0$ axis
which is the direction passing through the two galaxy centers.
The galaxy separation is $R_{12}=0.25~h^{-1}$ Mpc. At small
scales one can clearly see the holes around the galaxies
(which may be an unrealistic feature of halo models).
On larger scales $q_{ggm}$ drops off slowly.
\label{fig:Q-slice}}
\end{figure}

\begin{figure}
\begin{center}
\epsfxsize=12.0cm \hspace{-2cm} \epsfbox{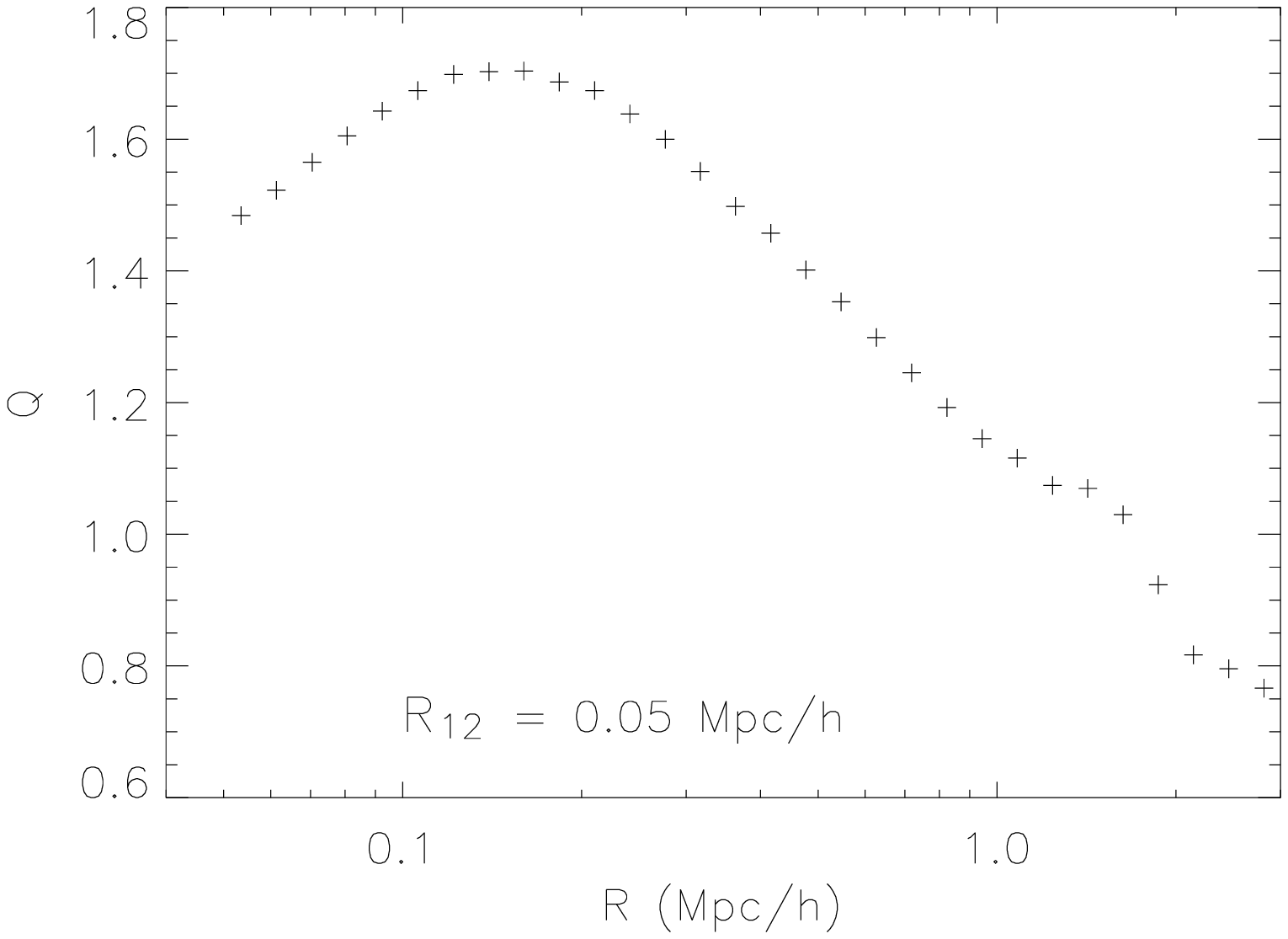}
\end{center}
\caption[The radial profile of the 2D $q_{ggm}$ for galaxies with
$R_{12}= 0.05~h^{-1}$ Mpc] {
The radial profile of the 2D $q_{ggm}$ for closer galaxy pairs,
$R_{12} = 50~h^{-1}$ kpc. The inner part of the curve
is still rising as $q_{ggm}$ climbs out of the holes.
At scales of a few times $R_{12}$ the profile is fairly
radially symmetric and decreases slowly. There is an apparent
bump at $1.5~h^{-1}$ Mpc where the two-halo term typically 
begins to dominate. 
\label{fig:Q-rad-profile}}
\end{figure}

Figure \ref{fig:Q-slice} shows the amplitude of $q_{ggm}$ along the $y=0$ axis,
now for galaxies separated by $R_{12}=0.25 ~h^{-1}$ Mpc. 
From this plot one can see that on larger scales $q_{ggm}$ drops
off slowly, but on small scales there are holes around the galaxy
pairs. On scales larger than the galaxy separation, $q_{ggm}$ is mostly
radially symmetric about the center. In Figure \ref{fig:Q-rad-profile}
we show the radial profile of $q_{ggm}$ about the center of two galaxies
with $R_{12} = 0.05 h^{-1}$ Mpc. On scales comparable to $R_{12}$,
$q_{ggm}$ is rising as it climbs out of the hole and at larger scales 
it drops off.

\section{Extracting Information from Shear Measurements}

As we saw in Section \ref{section:gal-gal-lensing}, we can use the method
of galaxy-galaxy lensing (GGL) to measure the galaxy-mass correlation function.
By measuring the tangential shear around lens galaxies, one can 
reconstruct $w_{gm}(R)$ and then invert the projection to obtain
$\xi_{gm}(r)$. In Section \ref{section:measuring-higher}, we calculated the
2D mass density around projected pairs of galaxies and showed how it was
related to the GGM3PCF. We will now discuss
how to measure this statistic from shear data using the SDSS as our fiducial
data set. We assume we have a spectroscopic
sample of galaxies with measured redshifts that we
will use as our lens galaxies and a fainter photometric sample of galaxies
without redshifts (though possibly with photometric redshifts)
that we will use as source galaxies.

Our first task is to find pairs of lens galaxies out to some maximum projected radius.
We will also want to impose a cut on their redshift difference. As we discussed
in Section \ref{section:measuring-higher}, this cut imposes an effective
cut in the radial direction $L_U$. One will probably want to keep
this fairly large, such as $30-50 ~h^{-1}$ Mpc, to limit the effects 
of redshift distortions on the interpretation of the results.

Once we have a galaxy pair, we rotate the local coordinate frame to
align them along the x-axis. There was no need to do this with GGL
since the average tangential shear around single galaxies 
must be isotropic. In that case, we just chose a set of radial
bins and averaged the tangential shear for 
objects in each bin. Now we have a pair
of galaxies which breaks this spherical symmetry and 
therefore need to deal with both components of the 
shear and keep track of both $x,y$ components
of the source galaxies. Thus, we now have a 2D problem and will
measure a 2D shear map rather than just a 1D shear profile. 
We choose an $x,y$ grid on which to store the ellipticity measurements
and use the rotated $x,y$ coordinates of the source galaxies to calculate
which pixel each source galaxy belongs to (for a given lens galaxy pair).
 
We want to build up a map of the average source galaxy ellipticity 
stacked (averaged) over many lens galaxy pairs. However, since the lens galaxy
pairs have a range of projected separations, we also need to bin in this
separation, $R_{12}$, as well and therefore produce a shear map for
each $R_{12}$ bin. If we bin too coarsely, this will have the effect of 
blurring together different $R_{12}$ correlation functions and 
we lose information. If we bin too finely, the number of lens pairs
per bin becomes smaller and the shot noise will overwhelm the signal.
With real data, one would also want to bin the len galaxy sample by
luminosity and perhaps by color, but we no not have that information
in our simulation. 

The fact that this is a 2D problem with no exact symmetries means that
instead of just a few radial bins one might have hundreds of pixels.
In addition, as noted above, one has to separate the lens galaxy 
pairs into several bins of $R_{12}$.
Because of this, one cannot expect a high signal-to-noise ratio in any
particular shear map pixel since the signal-to-noise ratio is only of order
1-10 in present measurements of the GGL signal from the GM2PCF with only a few radial bins.
However, as we will see, one can use approximate symmetries to
make measurements with signal-to-noise similar to the simpler GGL measurements
of the GM2PCF.

\begin{figure}
\begin{center}
\epsfxsize=12.0cm \hspace{-2cm} \epsfbox{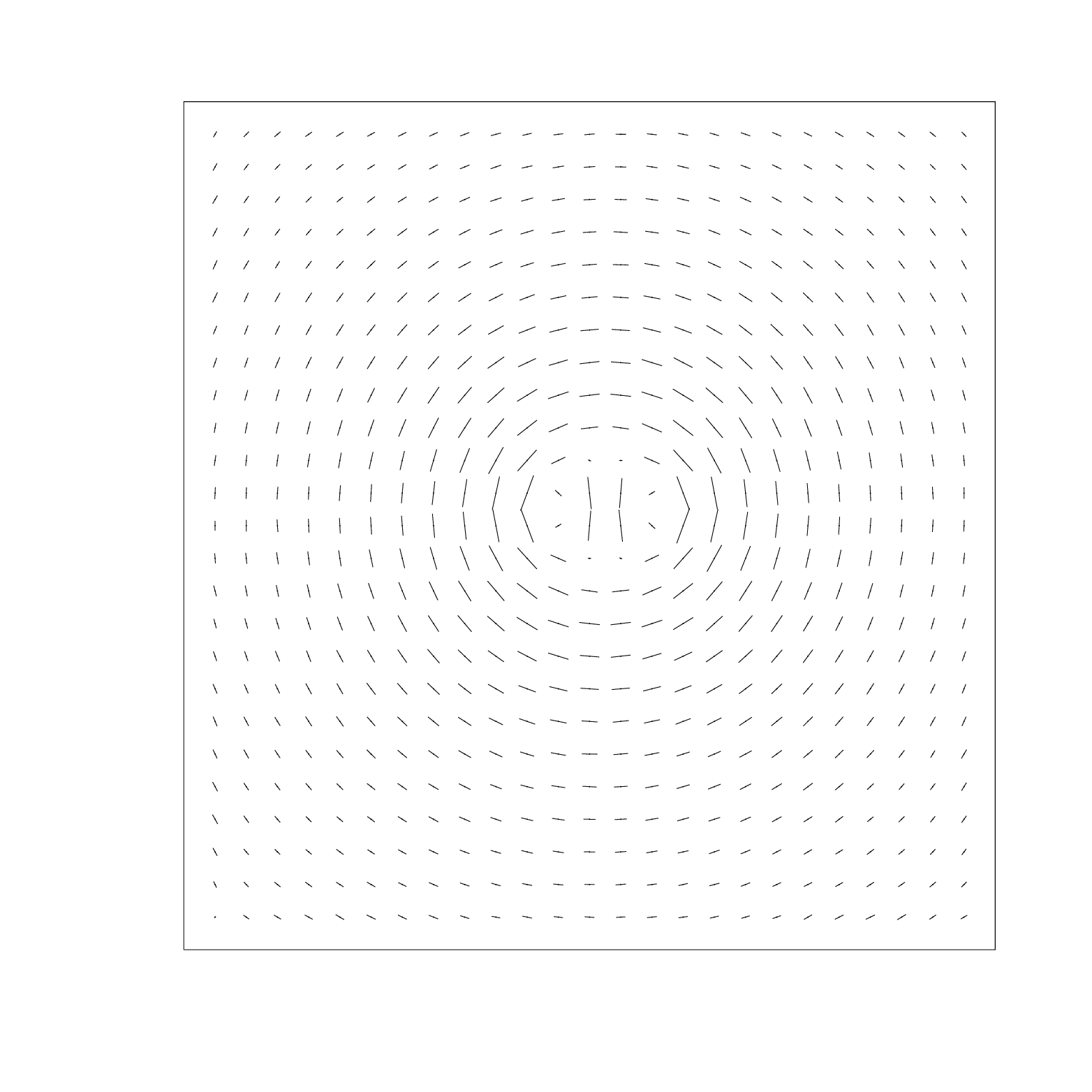}
\end{center}
\caption[The shear map around galaxies separated by 
$R_{12}= 0.8~h^{-1}$ Mpc] {
The mean shear map around pairs of galaxies of separation
$R_{12}=0.8~h^{-1}$ Mpc. The size of the region is $6~h^{-1}$ Mpc on 
a side. At larger scales the signal is mostly tangential but in the middle
the signal is more complicated. See next figure for a more detailed view. 
\label{fig:shear-map-whole}}
\end{figure}

\begin{figure}
\begin{center}
\epsfxsize=12.0cm \hspace{-2cm} \epsfbox{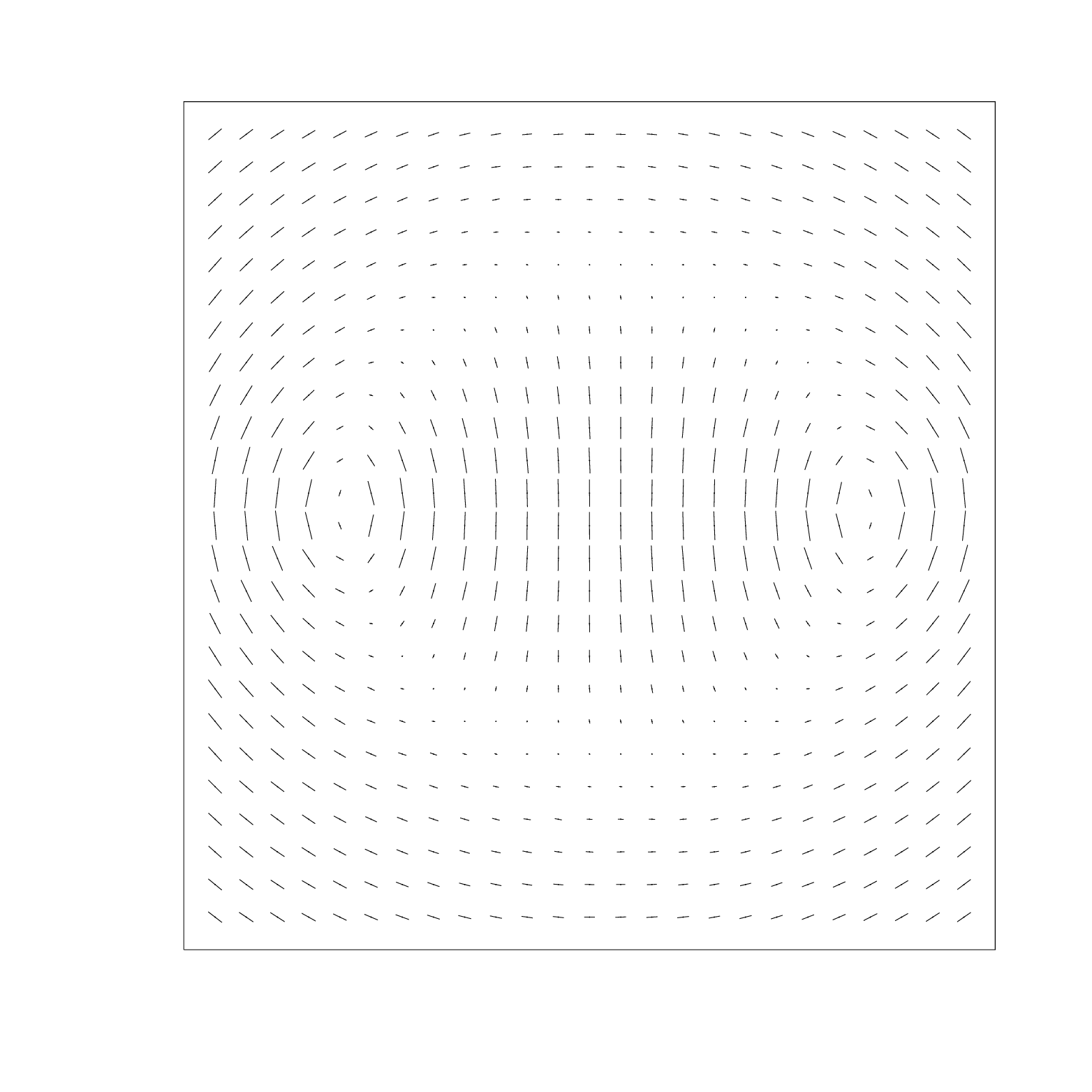}
\end{center}
\caption[A close up view of 
the shear map around galaxies separated by 
$R_{12}= 0.8~h^{-1}$ Mpc] {
A zoom of the previous figure of the central region near the
two galaxies. The scale is $1.2~h^{-1}$ Mpc on a side. 
In the middle, the mostly-tangential shear from each galaxy adds constructively.
There is also a ring of destructive interference where
the shear from each galaxy is in the opposite direction of the other. 
\label{fig:shear-map-close}}
\end{figure}

To calculate the shear map from $\Sigma(x,y)$ we solve for the 2D 
lensing potential $\phi(x,y)$ by solving Poisson's equation
(see Section \ref{section:weak-lensing-basics}). Then we numerically
differentiate $\phi$ to obtain the shear 
components $\gamma_1$ and $\gamma_2$. We will use as our model
the density $\Sigma(x,y)$ for $R_{12}=0.8~h^{-1}$ Mpc.
For simplicity, in calculating $\Sigma_{crit}$,
we choose all the lens redshifts 
to be $z_L=0.1$ and all the source
redshifts to be $z_S=0.3$, similar to the mean lens and source
redshifts used in the SDSS lensing measurements.
We display the shear map for a box of width $6~h^{-1}$ Mpc
calculated from this density in Figure \ref{fig:shear-map-whole}.
In Figure \ref{fig:shear-map-close} we show a higher resolution
image of the region close to the galaxy pair. This smaller region
is $1.2~h^{-1}$ Mpc on a side. From the first plot one can see that
at larger radius the shear is mostly tangential though one can notice
a residual ellipticity to the pattern that drops off as one goes out
further. 

In the central region, the shear map is more complicated and it helps
to look at the second plot, Figure \ref{fig:shear-map-close}. To
understand this pattern one has to remember that the shear is
a vector (technically a spin-2 field) and this pattern is the
superposition of two terms which are mostly tangential about the
two galaxies. Because it is a vector, there can be
constructive and destructive interference. The region in the
middle is a region of constructive interference where the
shear from each galaxy has the same phase (i.e., direction),
so this is where the shear is the largest. There
is also a ring of destructive interference where the shear from
each galaxy cancels out the other and no net shear is produced.

\begin{figure}
\begin{center}
\epsfxsize=12.0cm \hspace{-2cm} \epsfbox{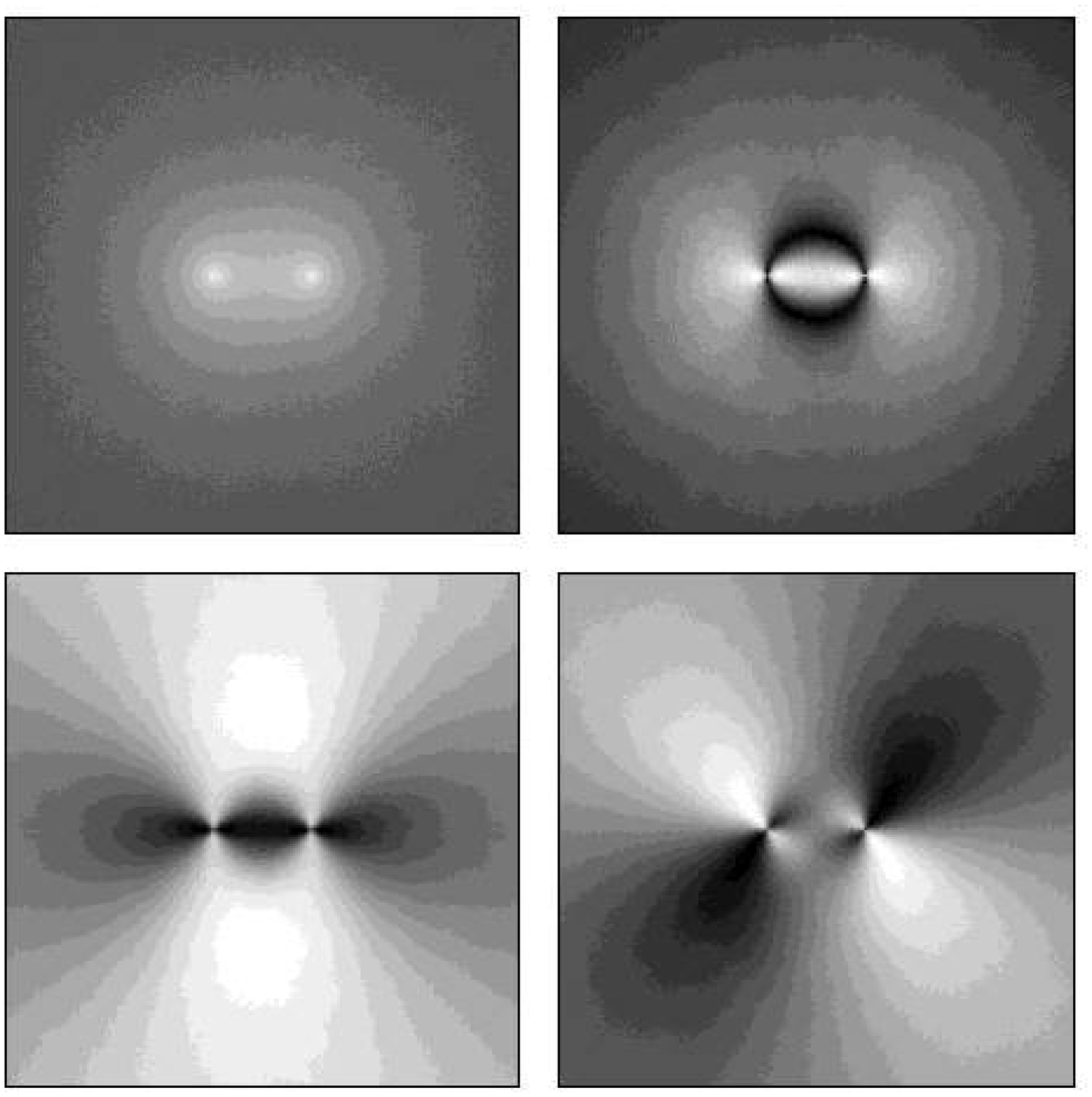}
\end{center}
\caption[The individual shear components 
for the map around galaxies separated by 
$R_{12}= 0.8~h^{-1}$ Mpc] {
Displayed are 
$\log \kappa$ (top left); $|\gamma|$ (top right); $\gamma_1$ (bottom left);
$\gamma_2$ (bottom right)
The galaxy separation is $R_{12}=0.8~h^{-1}$ Mpc
and the box size is $4.2~h^{-1}$ Mpc on a side.    
\label{fig:shear4}}
\end{figure}

In Figure \ref{fig:shear4} we show the individual shear components
around the same 2D density map. Here the scale is $4.2~h^{-1}$ Mpc on
a side. The top left shows $\kappa =\Sigma / \Sigma_{crit}$.
Top right shows the magnitude of the shear
$|\gamma| = \sqrt{\gamma_1^2 + \gamma_2^2}$. The bottom panels
show $\gamma_1$ and $\gamma_2$ from left to right respectively.

The $\kappa$ plot shows the bimodal pattern on small scales and elliptical
contours on larger scales. The plot of $|\gamma|$ shows a dark ring
where the net shear is zero due to the interference and the bright spots
where the shear is largest in the center and just on the outsides
of where the galaxies are. It also shows 
how $|\gamma|$ has elliptical contours on larger scales. 

\begin{figure}
\begin{center}
\epsfxsize=12.0cm \hspace{-2cm} \epsfbox{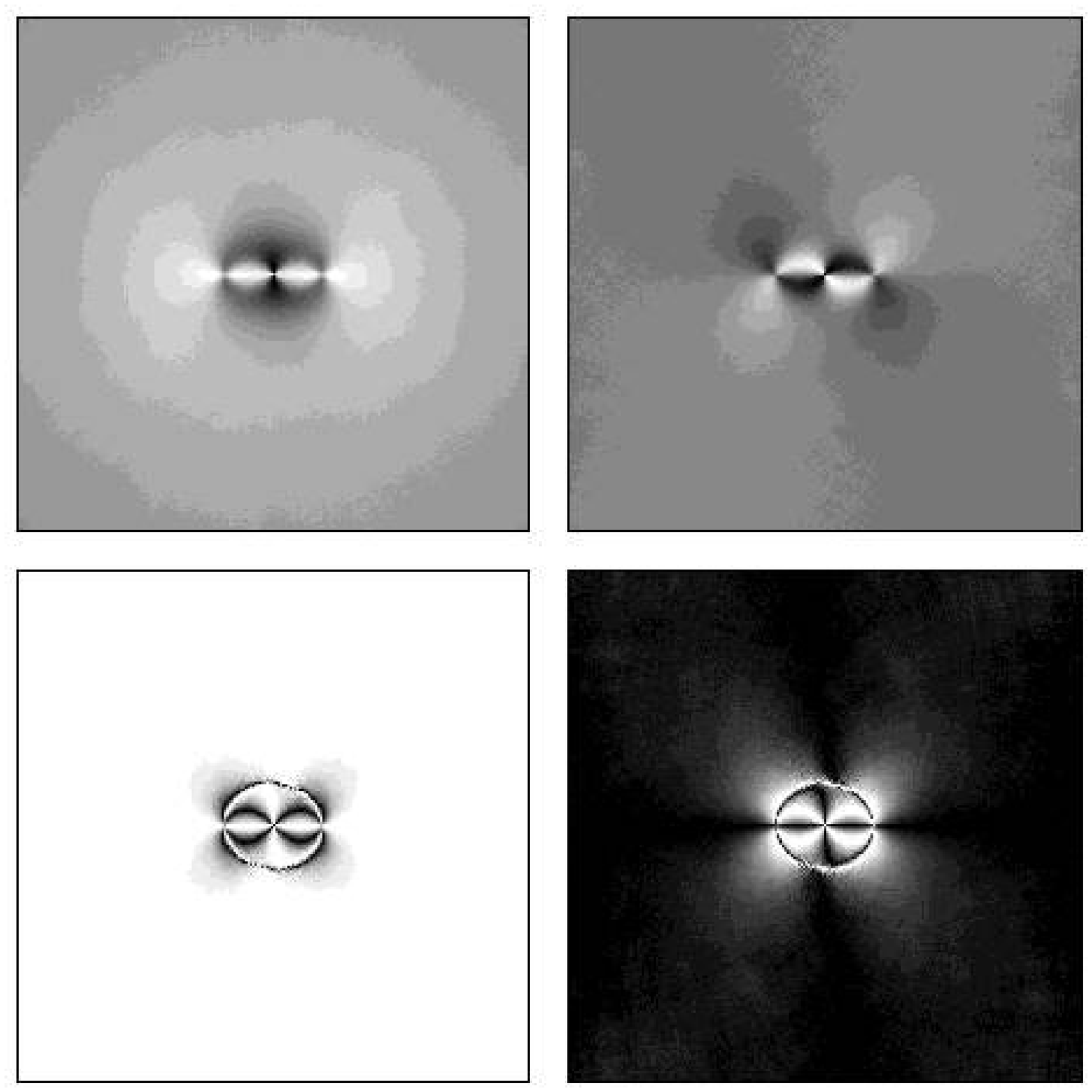}
\end{center}
\caption[The tangential and radial 
shear components for the map around galaxies separated by 
$R_{12}= 0.8~h^{-1}$ Mpc] {
Displayed are $\gamma_T$ (top left);
$\gamma_R$ (top right); $|\gamma_T|/|\gamma|$ (bottom left);
$|\gamma_R|/|\gamma|$ (bottom right);
The galaxy separation is $R_{12}=0.8~h^{-1}$ Mpc
and the box size is $4.2~h^{-1}$ Mpc on a side.    
\label{fig:shear4-tan}}
\end{figure}

In Figure \ref{fig:shear4-tan} we show the tangential shear
$\gamma_T$ and the radial shear $\gamma_R$ 
defined as
$\gamma_T \equiv -\gamma_1 \cos 2\theta - \gamma_2 \sin 2\theta$ and 
$\gamma_R \equiv -\gamma_2 \cos 2\theta + \gamma_1 \sin 2\theta$
where $\theta$ is the angle from the midpoint of the $x$-axis.
The top left is $\gamma_T$ and top right $\gamma_R$. The bottom
left is $|\gamma_T|/|\gamma|$ and bottom right is $|\gamma_T|/|\gamma|$.
In the bottom color scale, black is zero and white is 1. The bottom
plots show how at large scales the shear is almost entirely 
tangential whereas at small scales the pattern is complicated and
both $\gamma_T$ and $\gamma_R$ make contributions. 

\begin{figure}
\begin{center}
\epsfxsize=12.0cm \hspace{-2cm} \epsfbox{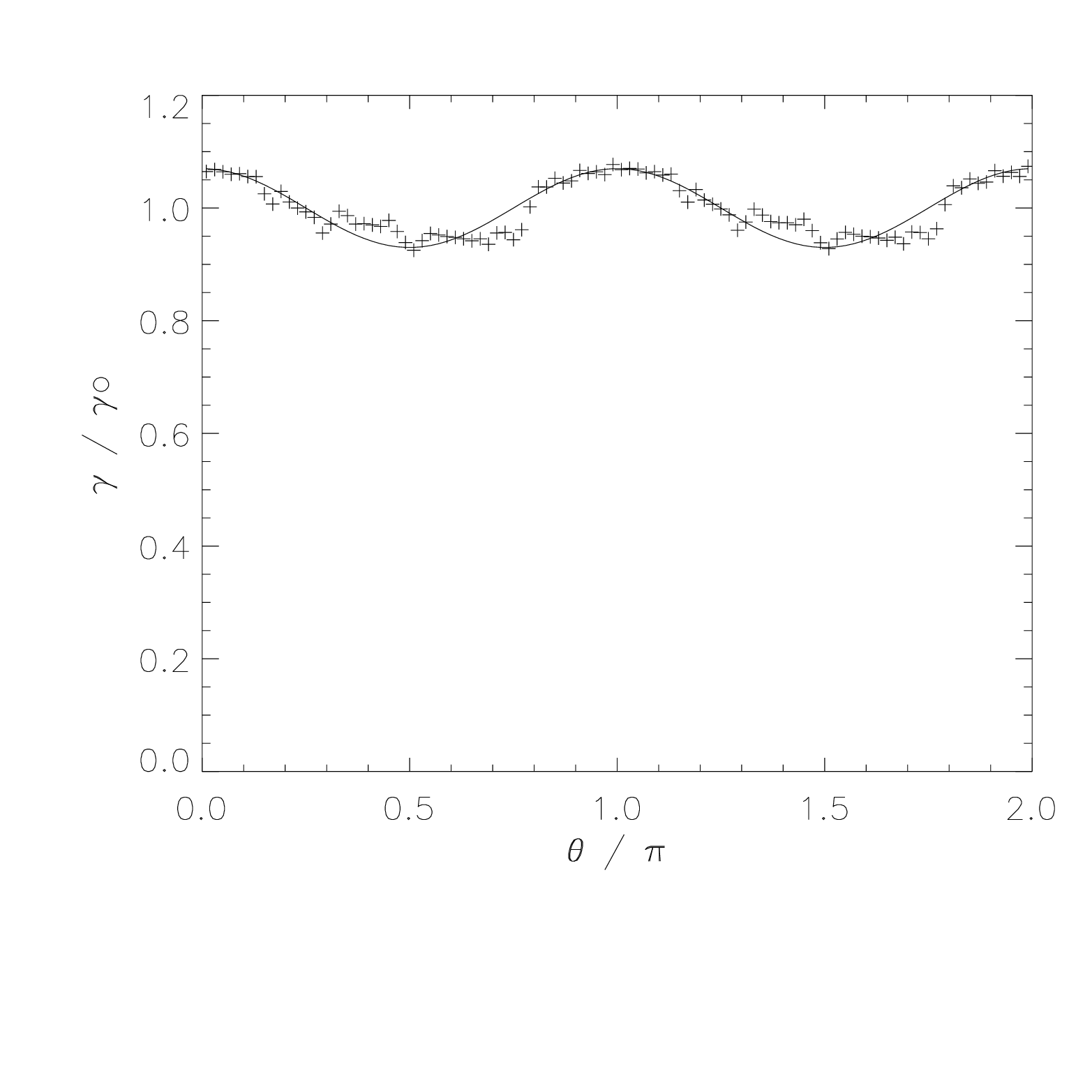}
\end{center}
\caption[The azimuthal variation in the tangential shear 
around galaxies pairs with $R_{12}= 2.0~h^{-1}$ Mpc] {
The tangential shear in an annulus 
of radius $R=2.0 \pm 0.1~h^{-1}$ Mpc, centered between lens galaxies
divided by the mean value in the annulus as a function of $\theta$.
A quadrupole pattern is evident and we over-plot a pure quadrupole
$\gamma_T/\gamma_{To}(\theta)=(1+\epsilon \cos 2\theta)$ with and
ellipticity parameter $\epsilon=0.07$.
\label{fig:shear-theta}}
\end{figure}

This is an example of how there is an approximate symmetry that
allows one to ignore the radial shear component at large scales.
On these scales one can just measure the tangential component and
see how its magnitude changes as a function of $R$ and $\theta$.
For example Figure \ref{fig:shear-theta} shows the tangential
shear in an annulus (centered between the galaxies) of radius
$R=2.0 \pm 0.1~h^{-1}$ Mpc, as a function of $\theta$
(measured from the $x$-axis), 
divided by the mean value in the annulus. A quadrupole pattern is 
evident and we over-plot the function 
$\gamma_T/\gamma_{To}(\theta)=(1+\epsilon \cos 2\theta)$ with
$\epsilon=0.07$. One can relate the moments of the shear map to
moments of $\kappa$ through methods developed by 
\cite{schneider-bartelmann:moments,natarajan:lens2d}.

\section{Estimate of signal-to-noise ratios}
\label{section:sig-noise}
So far we have calculated the shear signal we expect 
around foreground galaxy pairs, and now we 
investigate the amount of noise we expect. The noise in a 
weak lensing measurement is almost entirely due to shot 
noise from galaxy shapes,
although on the largest scales, sample variance can increase the 
variance and add some covariance \citep{sheldon:gmcf}. We will only
discuss the shot noise since it is the dominant source of noise 
on scales less than a few Mpc. The shot noise on a particular
measurement of $\gamma_i$ is about $0.2/\sqrt{N}$, where $0.2$ represents the
RMS intrinsic ellipticity of galaxies divided by 2. The number $N$ for
GGL is the number of lens-source galaxy pairs in a given radial bin.
For our analysis, we have pairs of lens galaxies and so $N$ is the number of 
lens-lens-source triplets in a given measurement bin. We now estimate 
what this number $N$ will be. 

If we have $N_p$ foreground galaxy pairs in a given projected 
separation bin, $R_{12}$, and a 2D number density $\sigma_S$ 
of source galaxies, then the number of triplets 
is just $N=N_p \sigma_S dA$ where $dA$ is 
some surface area element in which we estimate the shear. We have assumed
that the source density $\sigma_S$ is uncorrelated with the density of foreground
pairs, which should be correct if lens and source galaxies are at different redshifts.
To estimate the number of lens galaxy pairs, we calculate the average number of
neighbors per galaxy, $\langle n_p \rangle$, in a thin 
annulus about projected separation $R$ of width $\Delta R$.
This is just an integral over the galaxy-galaxy correlation function
\citep{peebles:lss},
\begin{eqnarray}
\langle n_p \rangle & = & n_g~2\pi~R~\Delta R ~\int_{-L_U/2}^{L_U/2}du~(1+\xi_{gg}(\sqrt{u^2+R^2})) \\
& = & n_g~2\pi~R~\Delta R ~(L_U+w_{gg}(R))
\end{eqnarray}
where $n_g$ is the 3D number density of lens galaxies. The total number of pairs 
in this bin is then
$N_p = N_L \langle n_p \rangle /2$, where $N_L$ is the total number of 
lens galaxies. The number $N_L$
is just the number density $n_g$ times the comoving volume 
probed by the survey, $V$. 
Finally we can write the number of lens-lens-source 
triplets, $N$, in surface area 
element $dA$ for lens pairs separated by $R \pm \Delta R/2$ as 
\begin{equation}
N= \pi~n_g^2~V~R~\Delta R~(L_U+ w_{gg}(R))~\sigma_S~dA
\end{equation}

The galaxy number density in our simulation is $n_g=4.56\times10^{-3}
h^{3} \mbox{Mpc}^{-3}$,
which roughly corresponds to an effective absolute magnitude cut of
$M_r < -20$. For this absolute magnitude cut, the SDSS probes an effective survey 
volume of approximately $V=1.2~f_{sky}~\times10^8 h^{-3} \mbox{Mpc}^{3}$ \citep{tegmark:pk},
where $f_{sky}$ is the fraction of the sky covered. For the SDSS, $f_{sky}$ will 
eventually be close to 1/4. The number density of source galaxies used for lensing
from \cite{sheldon:gmcf} is $3200/deg^2$. The area element dA in $deg^2$ is
dA=$(180~dx/\pi D(z_L))^2$ where $dx$ is the pixel size in physical units that
we will use for our shear map and $D(z_L)$ is the angular diameter distance
to redshift $z_L$. We will choose $z_L=0.1$, the median redshift
for the SDSS spectroscopic survey, and $7000~deg^2$ for the estimated 
sky coverage of the completed SDSS,
which gives $f_{sky}=0.17$. With these numbers our expression for $N$ becomes
\begin{equation}
N=2.0 \times 10^5 h^{5} Mpc^{-5}~ R~\Delta R ~(L_U + w_{gg}(R)) ~dx^2 
\end{equation}

\begin{figure}
\begin{center}
\epsfxsize=12.0cm \hspace{-2cm} \epsfbox
{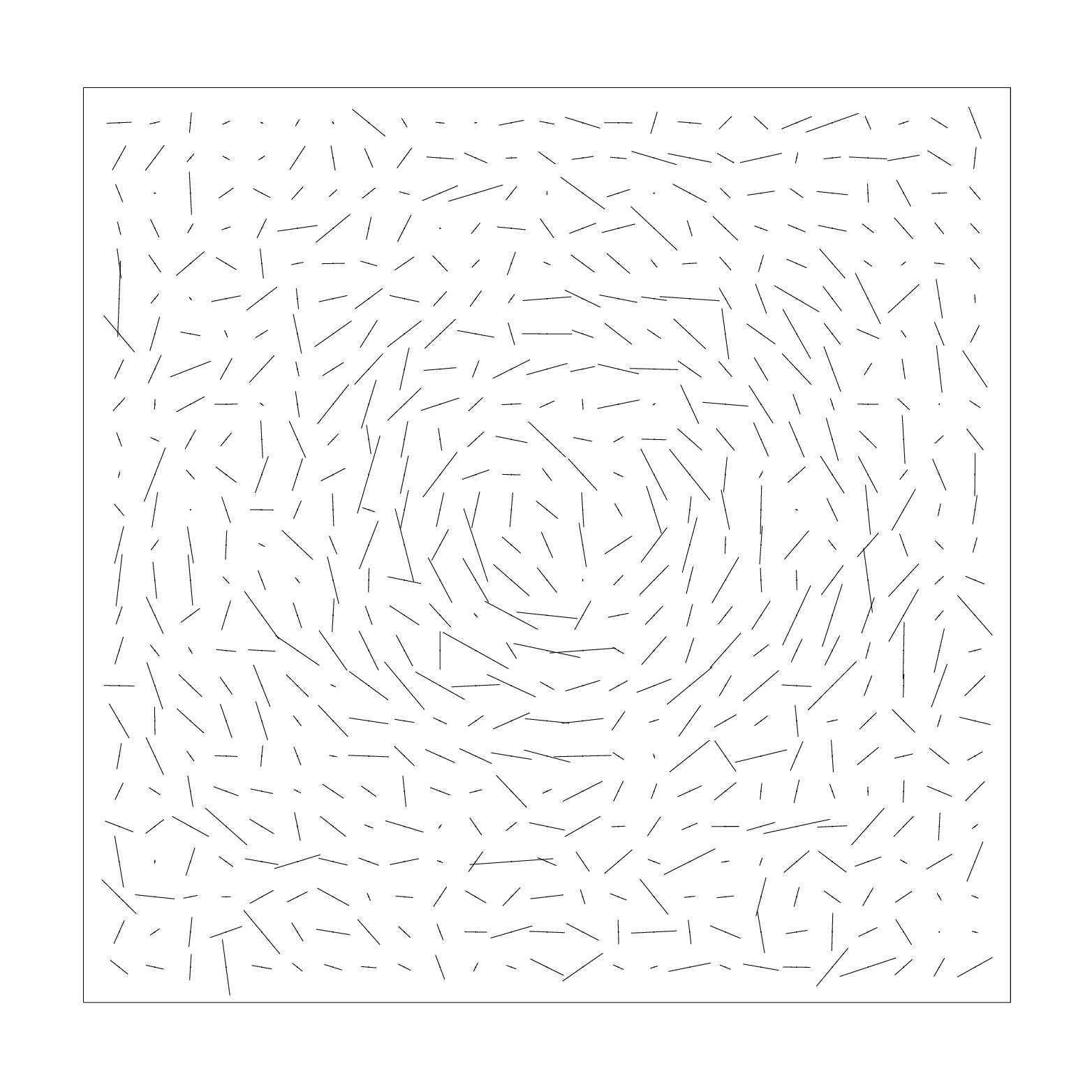}
\end{center}
\caption[The shear map around galaxies separated by 
$R_{12}= 0.8~h^{-1}$ Mpc with noise added] {
The same shear map of Figure \ref{fig:shear-map-whole}
when Gaussian noise, expected from the SDSS, is added. 
\label{fig:shear-map-noise}}
\end{figure}

With this formula we are able to add noise to our shear maps that should
correspond to what we will see if we do an analysis similar to 
\cite{sheldon:gmcf} with SDSS data in the near future to measure the
mass density around pairs. We take our predicted shear map shown in Figure 
\ref{fig:shear-map-whole}, which corresponds to $R_{12}=0.8~h^{-1}$ Mpc,
and add Gaussian noise with $\sigma=0.2/\sqrt{N}$ to each component of the shear. 
The resulting noisy shear map is shown in Figure \ref{fig:shear-map-noise}. The
basic tangential pattern can still be seen behind the noise.

We can now process both the noisy shear map and the noiseless one to
reconstruct $\kappa$ with a KS mass reconstruction algorithm. We use 
the ``direct method'' of \cite{lombardi-bertin:inversion}. This inversion
routine reconstructs the mass map by solving for the Fourier modes on
a chosen grid that minimize the functional
\begin{equation}
S=\frac{1}{2} \int_A d^2x || \nabla \kappa(\vec{x}) - \vec{u}(\vec{x})||
\end{equation}
where 
\begin{equation}
\vec{u}(\vec{x}) \equiv \left (  \begin{array}{c}
	\gamma_{1,1} + \gamma_{2,1} \\ \gamma_{2,1} - \gamma_{1,2} \label{eq:shear-u-def}
	\end{array} \right )    
\end{equation}
The Euler-Lagrange equations of this functional are just Equations
\ref{eq:kaiser-del} that we saw related the measured shear to $\kappa$.
This routine is linear and can be done quickly with Fast Fourier Transforms.
Unlike the earliest KS algorithms this one does not have edge effect biases
other than slightly increased noise in the bounding bins.
The process involves choosing a grid scale, averaging the shear in the grid
bins, calculating $\vec{u}(\vec{x})$ by finite differencing the shear values, 
and then running Lombardi and Bertin's routine (which is publicly available)
on the array $\vec{u}$. The resulting reconstruction will be degraded in 
resolution for two reasons. First, one will usually 
want to choose a grid scale that gives reasonable
signal-to-noise per pixel. This is purely a pixelization effect and structures
smaller than the pixel scale will be lost. Second, the reconstruction algorithm 
which differences neighbors introduces an effective smoothing between 
neighboring pixels. 

\begin{figure}
\begin{center}
\epsfxsize=12.0cm \hspace{-2cm} \epsfbox{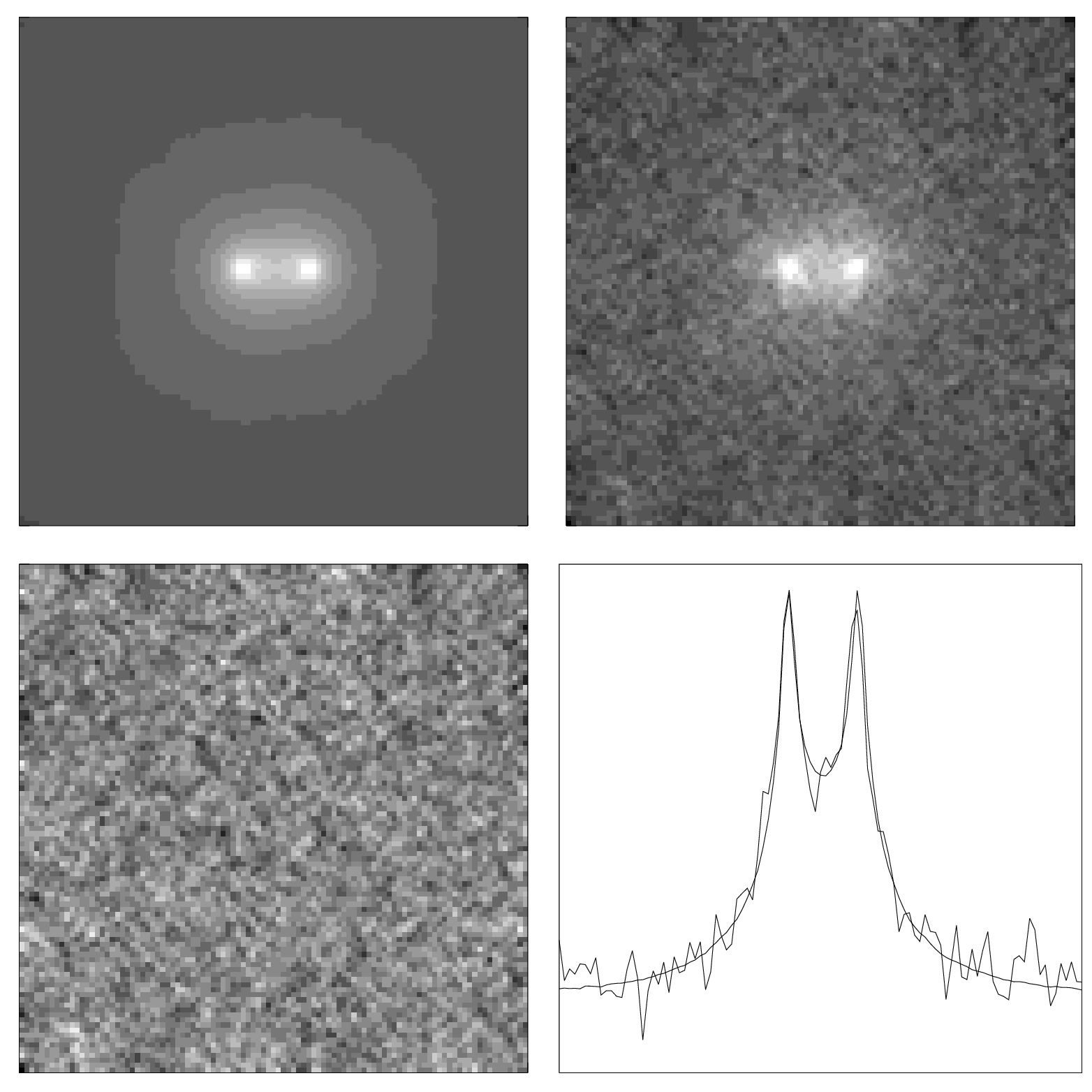}
\end{center}
\caption[The reconstruction of the mass map from the
noisy shear map] {
The top left shows the original $\kappa$ map binned down
to the same resolution of the reconstructed one. The top
right shows the reconstruction obtained by inverting the
noisy shear map. The bottom left is the residual from
subtracting the top two maps. One can see that the residual
is consistent with zero but the noise is also correlated
over the scale of just a couple of pixels. The bottom right
plot shows a slice through the two peaks for both densities.
One can see that the reconstruction works very well.
The means have been subtracted off both maps before plotting
since this constant mass sheet is unmeasurable.
\label{fig:recon4}}
\end{figure}

We display the reconstruction in Figure \ref{fig:recon4}. The top right is the
original $\kappa$ binned down to the same grid resolution of the
reconstruction. The top right shows the $\kappa$ recovered from the 
noisy shear map via the reconstruction algorithm. The bottom left shows the
residual, which is consistent with zero. The bottom right shows a slice through 
both maps that intersects both peaks around the galaxies. Since the 
reconstruction cannot recover the constant mass sheet we have subtracted off
the mean from each map before plotting.

\section{Conclusion}

We have calculated the weak lensing shear around pairs of galaxies and 
related it directly to the 3D galaxy-galaxy-mass three-point correlation function.
We have shown that the 2D mass profile can be reconstructed from the shear
map around galaxy pairs and we have discussed how to interpret this in terms
of the correlation functions. With our N-body simulation and HOD bias model we 
have calculated the expected signal and have estimated the amount of noise to
be expected in a model survey similar to the Sloan Digital Sky Survey. 
Although this prediction is only meant to be a first attempt at estimating the 
effect and does not exhaust the possibilities we predict that this 
signal may be measurable with current SDSS lensing data. The anisotropic dark 
matter profile about two galaxies extends far away from their center and will 
produce an anisotropic tangential shear profile. Measurements confirming this will
be difficult to reconcile with alternative theories of gravity that need to
produce weak lensing shear without any extended anisotropic dark matter halos.
On the other hand we have seen that they are a natural prediction of CDM.
The three-point function better specifies how mass traces galaxies and
so provides more detailed information over that contained in the
isotropic galaxy-mass two-point correlation function. This exact form 
of the measured galaxy-galaxy-mass correlation function will provide 
additional constraints on galaxy bias models that govern how galaxies
trace the underlying dark matter.

\section{Acknowledgments}
I would like to thank Joshua Frieman, Erin
Sheldon and Andreas Berlind for useful discussions. I would also like
to thank Martin White for providing the N-body simulations 
and Andreas for providing the galaxy occupation. I would also like to thank the
Aspen Center for Physics where some of this work was completed.

\end{document}